\documentclass[sigconf]{acmart}

\usepackage{xcolor}
\usepackage[most]{tcolorbox}
\newtcolorbox{mybox}[2][]{%
  attach boxed title to top center
               = {yshift=-8pt},
  colback      = gray!5!white,
  colframe     = gray!75!black,
  fonttitle    = \bfseries,
  colbacktitle = gray!85!black,
  left=0pt,
right=0pt,
top=0pt,
bottom=0pt,
  title        = #2,#1,
  enhanced,
}

\usepackage[english]{babel} 
\usepackage[T1]{fontenc}
\usepackage{graphicx}

\usepackage{subfig}

\usepackage{gensymb}
\definecolor{myBlue}{RGB}{0,133,255}%{HTML}{0085ff}
\usepackage{tikz}
\newcommand\phase[1]{\tikz[baseline=(X.base)]\node [draw=myBlue,fill=myBlue,thick,rectangle,inner sep=2pt, rounded corners=2pt](X){\color{white}\textbf{#1}};}

\usepackage{multicol}

\DeclareMathAlphabet{\mathrmbf}{\encodingdefault}{\rmdefault}{bx}{n}
\DeclareMathAlphabet      {\mathbfit}{OML}{cmm}{b}{it}

\usepackage{booktabs} % For formal tables
\usepackage{amssymb}
\usepackage{array}
\usepackage{ltl}
\usepackage{thmtools} 
\usepackage{thm-restate}
\acmConference[]{}{}{}
\usepackage[xcolor=orange]{changes}
\definechangesauthor[name={Claudio Menghi},color=orange]{CM}
\definechangesauthor[name={Shiva Nejati},color=red]{SN}
\usepackage[colorinlistoftodos,prependcaption,textsize=tiny]{todonotes}

\newboolean{showcomments}
\setboolean{showcomments}{true} % toggle to show or hide comments
\ifthenelse{\boolean{showcomments}}
{\newcommand{\nb}[2]{
  \fcolorbox{black}{yellow}{\bfseries\sffamily\scriptsize#1}
  {\sf\small$\blacktriangleright$\textit{#2}$\blacktriangleleft$}
 }
 
}
{\newcommand{\nb}[2]{}
 
}

 \newcommand{\goal}[1]{\textbf{\textsc{O#1}}} 
 \newcommand{\assumption}[1]{\textbf{\textsc{A#1.}}}

\usepackage{pifont}% http://ctan.org/pkg/pifont
\newcommand{\cmark}{\ding{51}}%
\newcommand{\xmark}{\ding{55}}%

\newcommand{\partner}{LuxSpace}

\newcommand\inputs{\ensuremath{I}}
\newcommand\inputsignal{\ensuremath{i}}

\newcommand\outputs{\ensuremath{O}}
\newcommand\outputsignal{\ensuremath{o}}

\newcommand\diff{\emph{diff}}
\newcommand\ourlogic{RFOL}
\newcommand\casestudy{\textsf{SatEx}}

\newcommand\system{M}

\makeatletter
\def\old@comma{,}
\catcode`\,=13
\def,{%
  \ifmmode%
    \old@comma\discretionary{}{}{}%
  \else%
    \old@comma%
  \fi%
}
\makeatother

\usepackage[inline]{enumitem}
\newcommand\interpretation[1]{\ensuremath{\llbracket #1 \rrbracket}}

\newcommand\timedomain{\ensuremath{\mathbb{T}}}
\newcommand\real{\ensuremath{\mathbb{R}}}

\newcommand\formula{\ensuremath{\varphi}}

\newcommand\testoracle{\ensuremath{\system_\formula}}

\usepackage{fancybox}

\usepackage[utf8]{inputenc}

\newcommand\numcasestudy{11}

\usepackage{numprint}
\usepackage{amsmath}
\npthousandsep{\,}

\setcopyright{none}
\renewcommand\footnotetextcopyrightpermission[1]{} 

\renewcommand\footnotetextcopyrightpermission[1]{} % removes footnote with conference information in first column
\begin{document}
\title[Generating Automated and Online Test Oracles for Simulink Models with Continuous and Uncertain Behaviors]{Generating Automated and Online Test Oracles for Simulink Models with Continuous and Uncertain Behaviors}

\author{Claudio Menghi}
\affiliation{
  \institution{SnT - University of Luxembourg}
}
\email{claudio.menghi@uni.lu}

\author{Shiva Nejati}
\affiliation{
  \institution{SnT - University of Luxembourg}
}
\email{shiva.nejati@uni.lu}

\author{Khouloud Gaaloul}
\affiliation{
  \institution{SnT - University of Luxembourg}
}
\email{khouloud.gaaloul@uni.lu}

\author{Lionel Briand}
\affiliation{
  \institution{SnT - University of Luxembourg}
}
\email{lionel.briand@uni.lu}

\begin{abstract}
Test automation requires automated oracles to assess  test outputs. 
For cyber physical systems (CPS), oracles, in addition to be automated, should ensure some key objectives: 
(i) they should check test outputs in an online manner  to stop expensive  test executions as soon as a failure is detected; (ii) they should handle time- and magnitude-continuous CPS behaviors;  
(iii) they should provide a quantitative degree of satisfaction or failure measure instead of binary pass/fail outputs; and 
(iv) they should be able to handle uncertainties  due to CPS interactions with the environment.
We propose an automated approach to  translate CPS requirements specified in a logic-based language into test oracles specified in Simulink -- a widely-used development and simulation language for CPS. 
 Our approach achieves the objectives noted above through the identification of a fragment of Signal First Order logic (SFOL) to specify requirements, the definition of a quantitative semantics for this fragment and a sound translation of the fragment into Simulink.  The results from applying our approach on \numcasestudy\ industrial case studies show that: (i)~our  requirements language  can express all the 98 requirements of our case studies; (ii) the time and effort required by our approach are acceptable, showing potentials for the adoption of our work in practice,  and (iii) for large models, our approach can dramatically reduce the test execution time compared to  
 when test outputs are checked in an offline manner. 
\end{abstract}

\maketitle

\section{Introduction}
\label{sec:intro}
The development of Cyber Physical Systems (CPSs) starts by specifying CPS control and dynamic behaviors as executable models described in languages such as Matlab/Simulink~\cite{mathworks}.  These models are complex and subject to extensive testing before they can be used as a basis for software code development.  Existing research on automated testing of CPS models has largely focused on automated generation of test suites~\cite{nardisurvey,harman2013comprehensive,oliveira2014automated}. However, in addition to test input generation, test automation requires \emph{automated oracles}~\cite{barr2015oracle}, i.e., a mechanism to automatically determine whether a test has passed or failed.

To automate oracles,  engineers  often rely  on runtime crashes (a.k.a. implicit oracles~\cite{pezze2014automated}) to detect failures.
However,  implicit oracles often cannot effectively reveal  violations of functional requirements as most of such violations do not lead to crashes.
As mandated by  safety certification standards~\cite{IEC61508}, for CPS,  functional requirements must be specified, and be used as the main authoritative reference to derive test cases and  to demonstrate system behavior correctness. To achieve this goal, we need to develop oracles that can automatically check the correctness of system behaviors with respect to requirements. In this paper, we  propose an approach to  generating oracles that \emph{automatically}  determine whether  outputs  of CPS models satisfy or violate their requirements.  For CPS, oracles, in addition to be automated, need to contend with a number of  considerations that we discuss and illustrate  below.

\begin{table*}[ht]
\caption{Requirements for the satellite control system (\casestudy) developed by \partner.}
\label{tab:requirements}
\scalebox{.8}{\begin{tabular}{l p{10cm} | p{8.5cm}}
\toprule
\textbf{ID} & \textbf{Requirement} & \textbf{Restricted Signal First-Order logic formula*} \\
\toprule
{\bf R1} & The angular velocity of the satellite shall always be lower than $1.5m/s$. & $\forall t \in [0, \numprint{86400}) \colon \|\vec{\mathit{w}}_{\mathit{sat}}(t)\|<1.5 $\\
\midrule
{\bf R2} & The estimated attitude of the satellite shall be always equal to 1. & $\forall t \in [0, \numprint{86400}) \colon \|\vec{q}_{estimate}(t)\| = 1 $\\
\midrule
{\bf R3} & The maximum reaction torque must be equal to $0.015Nm$. & $\forall t \in [0, \numprint{86400}) \colon 
\|\vec{trq}(t) \| \leq 0.015  $ \\
\midrule
{\bf R4} & The satellite attitude shall reach close to its target value within $\numprint{2000}$~sec (with a deviation not more than $2$ degrees) and remain close to its target value. & $\forall t \in [\numprint{2000}, \numprint{86400})  \colon  \| \vec{q}_{real}(t)-\vec{q}_{target}(t) \| \leq 2 $\\
\midrule
{\bf R5} & 
The  satellite target attitude shall not change abruptly:
for every $t$, the difference between the current target attitude  and the one at two seconds later shall not be more  than $\alpha \degree$. 
& $\forall t \in [0, \numprint{86400})  \colon  \| \vec{q}_{target}(t)-\vec{q}_{target}(t+2) \| \leq 2 \times sin(\frac{\alpha}{2})$\\
\midrule 
  {\bf R6} &  The satellite shall reach close to its desired attitude (with a deviation not more than \%2) exactly $2000$~sec after it enters its normal mode (i.e., $sm(t)=1$) and after it has stayed in that mode for at least 1~sec. 
   &  
   $\forall t \in [0, \numprint{86400}) 
   \colon (sm(t)=0 \wedge 
   (\forall t_1 \in (t, t+1] \colon sm(t_1)=1) \rightarrow \newline                                                                    
  \| \vec{q}_{real}(t+2000)-\vec{q}_{estimate}(t+2000) \| \leq 0.02)$\\ 
\bottomrule
\end{tabular}} 
\flushleft{\hspace*{.7cm}\small * The notation $\vec{a}$  indicates that $a$ is a vector; $\| \vec{a} \|$ indicates the norm of the vector.}
\vspace{-0.3cm}
\end{table*}

\textbf{Motivating Example.} We motivate our work using   \casestudy, a real-world case study from the satellite domain~\cite{Luxspace}. \casestudy\ is a  model of a satellite control system modeled in  
the Matlab/Simulink language and developed by our  partner \partner. Briefly, the main functions of the \casestudy\ model are:  (i)~to always keep the satellite on the desired orbit, 
(ii)~to ensure that the satellite is always facing the earth (i.e., the satellite's attitude is always pointing to the earth),  and (iii)~to regulate the satellite speed. The main functional  requirements of \casestudy\ are presented in the middle column of Table~\ref{tab:requirements}, and  the variables used in the  requirements are described in Table~\ref{tab:variables}.

Before software coding or generating code from Simulink models (a common practice when Simulink/Matlab models are used), engineers need to ensure that their models satisfy the requirements of interest (e.g., those in Table~\ref{tab:requirements}). Although there are a few automated verification  tools  for Simulink, in practice, verification of CPS Simulink models largely relies on simulation and testing.  This is because existing tools for verifying Simulink models~\cite{qvtrace,roy2011simcheck,SDV,REACTSYS} are not amenable to verification of large Simulink models like \casestudy\ that contain continuous physical computations  and third-party library code~\cite{matinnejad2018test,Abbas:13}. Further, CPS Simulink models often capture dynamic and hybrid systems~\cite{alur:15}. It is well-known that model checking such systems is in general undecidable~\cite{henzinger1998s,alur1995algorithmic,6064535}.

\begin{table}
\caption{Signals variables of the \casestudy\ model.}
\label{tab:variables}
\scalebox{.8}{\begin{tabular}{l l l l }
\toprule
\textbf{Var.} & \textbf{Description} & \textbf{Var.} & \textbf{Description}\\
\toprule
$sm$ & Satellite mode status. & $\vec{\mathit{trq}}$  &  Satellite torque. \\
\midrule
$\vec{\mathit{w}}_{\mathit{sat}}$ &  Satellite angular velocity. & $\vec{q}_{\mathit{real}}$ &  Current satellite attitude. \\
\midrule
$\vec{q}_{\mathit{estimate}}$ & Estimated satellite attitude. &$\vec{q}_{\mathit{target}}$ & Target satellite attitude. \\
 \bottomrule
\end{tabular}}
\end{table}

To effectively test CPS models, engineers need to have automated test oracles  that can check the correctness of simulation outputs with respect to the requirements. To be effective in the context of CPS testing, oracles should further ensure the following objectives:  

\begin{figure}[t]
\includegraphics[width=\columnwidth]{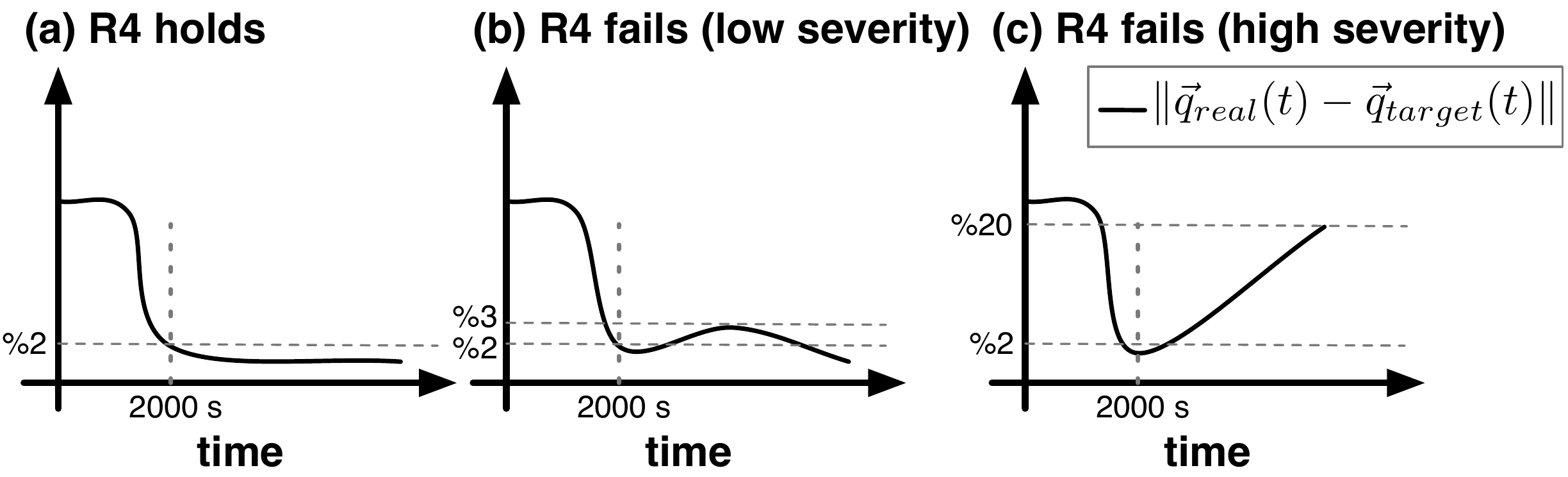}
\caption{Three simulation outputs of our \casestudy\ case study model indicating the error signal $\mathbf{ \|} \vec{\mathbf{q}}_{\mathbf{real}}\mathbf{(t)}-\vec{\mathbf{q}}_{\mathbf{target}}\mathbf{(t)} \mathbf{\|}$. The signal in~(a) passes  {\bf R4} in Table~\ref{tab:requirements}, but those in~(b) and (c) violate {\bf R4} with low and high severity, respectively.}
 \label{fig:motivation}
 \vspace{-0.3cm}
 \end{figure}

\goal{1}.
\emph{Test oracles should check outputs in an  online mode.}
An online oracle (a.k.a as a monitor in the literature~\cite{Bakhirkin:2018:FLS:3283535.3283536}) checks output signals as they are generated by the model under test. Provided with an online oracle, engineers can stop model simulations  as soon as failures are identified. Note that CPS Simulink models are often computationally expensive because they have to capture physical systems and processes  using high-fidelity mathematical models with continuous behaviors. Further, CPS models have to be executed for a large number of test cases. Also, due to the reactive and dynamic nature of CPS models, individual test executions (i.e., simulations) have to run for long durations  to exercise  interactions between the system and the environment over time.   For example, to simulate the satellite behavior for $24h$ (i.e., $\numprint{86400}$s), the \casestudy\ model has to be executed for $84$ minutes (\char`\~1.5 hours) on $12$-core Intel Core $i7$ $3.20$GHz $32$GB of RAM. Further, the $24h$-length simulation of \casestudy\ has  to be (re)run for  tens or hundreds of test cases.  Therefore, online test oracles are instrumental to reduce the total test execution time and to increase the number of  executed test cases within a given test budget time.

\goal{2}.
\emph{Test oracles should be able to evaluate  time and magnitude-continuous signals.}
CPS model inputs and outputs are  signals, i.e., functions over time. Signals are classified based on their time-domain into time-discrete and time-continuous, and based on their value-range into magnitude-discrete and magnitude-continuous. The type of input and output signals depends on the modeling formalisms. For example, differential equations~\cite{newton1774methodus} often used in physical modeling yield continuous signals, while finite state automata~\cite{FSA} used to specify discrete-event systems generate discrete signals. Figure~\ref{fig:motivation} shows three magnitude- and time-continuous signal outputs of \casestudy\ indicating the error in the  satellite attitude, i.e.,  the difference between the real and the target satellite attitudes  ($  \| \vec{q}_{real}(t)-\vec{q}_{target}(t) \| $).  An effective CPS testing framework should be able to handle the input and output signals of different CPS formalisms including the most generic and expressive signal type, i.e., time-continuous and magnitude-continuous. Such testing frameworks are then able to handle any discrete signal as well.

\goal{3}.
\emph{Test oracles for CPS  should provide a quantitative measure of the degree of satisfaction or violation of a requirement.} Test oracles typically classify test results as failing and passing. The boolean partition into ``pass" and ``fail", however,  falls short of the practical needs. In the CPS domain,  test oracles  should assess test results in a more nuanced way to identify among all the passing test cases, those that are more acceptable, and among all the failing test cases, those that reveal more severe failures.  Therefore,  an effective test oracle for CPS  should  assess test results using  a \emph{quantitative fitness measure}. For example, the satellite attitude error signal in Figure~\ref{fig:motivation}(a) satisfies the requirement {\bf R4} in Table~\ref{tab:requirements}. But, signals in Figures~\ref{fig:motivation}(b) and (c) violate  {\bf R4} since the error signal  does not remain below the $\%2$ threshold after $2000$s. However, the failure in  Figure~\ref{fig:motivation}(c) is more severe than the one in Figure~\ref{fig:motivation}(b) since the former deviates from the threshold with a larger margin. A quantitative oracle can differentiate between these failure cases and assess their degree of severity.

\goal{4}.
\emph{Test oracles should be able to handle uncertainties in CPS function models.}
There are various sources of uncertainty in CPS~\cite{moiz2017uncertainty}. In this paper, we consider two main recurring and common sources of uncertainties in CPS~\cite{elbaum2014known,de2007classification}: (1)~Uncertainty due to unknown hardware choices which results in model parameters whose values are only known approximately at early  design stages. For example, in \casestudy,  there are uncertainties in the type of the magnetometer and in the accuracy of the sun sensors mounted on the satellite (see Table~\ref{tab:uncertainParameters}).  (2)~Uncertainty due to the noise in the  inputs received from the environment, particularly in the sensor readings. This is  typically captured by  white noise signals applied to the model inputs (e.g.,  Table~\ref{tab:uncertainParameters} shows the signal-to-noise (S2N) ratios for the magnetometer and sun sensor inputs of \casestudy). Oracles for CPS  models should be able to assess outputs of models that contain parameters with uncertain values and signal inputs with noises.  
 
 \begin{table}
\caption{Uncertainty in \casestudy: 
The values of the magnetometer type and the sun sensor accuracy parameters are given as ranges (middle column). The  noise values for the magnetometer and sun sensor inputs are given in the right column.}
\label{tab:uncertainParameters}
\scalebox{.85}{
\begin{tabular}{l | l | l l} 
\toprule
\textbf{Component} & \textbf{Parameter Values} & \textbf{Noises (S2N)} \\
\toprule
Magnetometer &  $[60000,140000]$ nT &  $100 \cdot e^{-12}$ T/$\sqrt{\text{Hz}}$ \\
\hline
Sun sensor & $2.9 \cdot 10^{-3} \pm 10\%$ &  $2.688 \cdot e^{-6}$ A \\
\bottomrule
\end{tabular}}
\end{table}

\textbf{Contributions.}  We propose  \emph{Simulink Oracles for CPS RequiremenTs with uncErtainty (SOCRaTEs)}, an approach for generating online oracles in the form of Simulink blocks based on CPS functional requirements (Section~\ref{sec:assumption}). Our oracle generation approach 
achieves the four objectives discussed above  through the following novel elements:

{\bf -} We propose Restricted Signals First-Order Logic (RFOL),  a signal-based logic language to specify CPS requirements (Section~\ref{sec:problemDef}). RFOL  is a restriction of Signal First Order logic~\cite{bakhirkin2018first} (SFOL) that can capture properties of time- and magnitude-continuous signals while enabling the generation of efficient, online test oracles. We define a quantitative 
semantics for RFOL to compute a measure of fitness for test results as oracle outputs. 
 
{\bf -} We develop  a procedure to translate RFOL requirements into automated oracles modeled in the Simulink language (Section~\ref{sec:contribution}). We prove the soundness of our translation with respect to the quantitative semantics of RFOL. Further,  we  demonstrate that: (1) the generated oracles  are able to identify failures as soon as they are revealed (i.e., our oracles are online); and (2) our oracles can handle models containing parameters with uncertain values and signal inputs with noises.  Finally, we have implemented our automated oracle generation procedure in a tool which is available online~\cite{AdditionalMaterial}.

We apply our approach to  \numcasestudy\  industry Simulink models from two  companies in the CPS domain.  Our  results show that our proposed logic-based requirements language (RFOL) is sufficiently expressive to specify all the 98 CPS requirements in our industrial case studies. Further, our automated translation can generate online test oracles in Simulink efficiently, and the effort of developing RFOL requirements is acceptable, showing potentials for the practical adoption of our approach. Finally, for large and computationally intensive industry models, our online oracles can bring about dramatic time savings  by stopping test executions  long before their completion when they find a failure, without  imposing a large time overhead when they run together with the model.

\textbf{Structure.} Section~\ref{sec:assumption} outlines SOCRaTEs and its underlying assumptions.
Section~\ref{sec:problemDef} presents the Restricted Signals First-Order Logic and its semantics.
Section~\ref{sec:contribution} describes our automated oracle generation procedure.
Section~\ref{sec:evaluation} evaluates  SOCRaTEs.
Section~\ref{sec:related} presents the related work and Section~\ref{sec:conclusion} concludes the paper.

\section{SOCRaTEs}
\label{sec:assumption}

\begin{figure}[t]
\includegraphics[width=1\columnwidth]{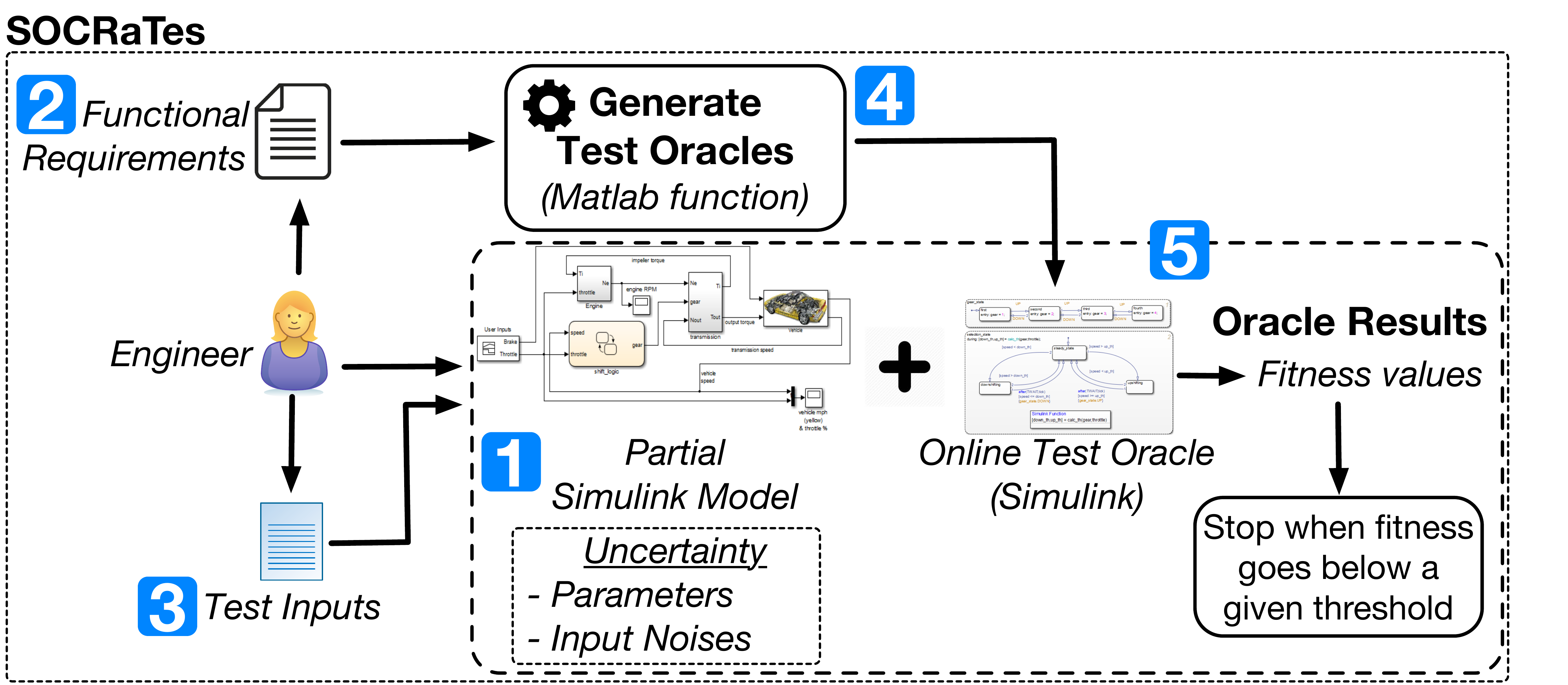}
\caption{Overview of SOCRaTes, our automated oracle generation approach.}
\label{fig:approach}
\end{figure}

Figure~\ref{fig:approach} shows an overview of SOCRaTeS~\cite{SOCRaTEs} (\emph{Simulink Oracles for CPS RequiremenTs with uncErtainty}),  our approach to generate automated test oracles for CPS models.  SOCRaTeS takes three inputs: (\phase{1}) a CPS model with parameters or inputs involving uncertainties, (\phase{2}) a set of functional requirements for the CPS model and (\phase{3}) a set of test inputs that are developed by engineers to test the CPS model with respects to its requirements. 
SOCRaTeS makes  the following assumptions about its inputs:

\assumption{1}
\emph{The CPS model is described in Simulink} (\phase{1}).
Simulink is used by more than 60\% of engineers for simulation of CPS~\cite{zheng2017perceptions,baresi2017test}, and is the prevalent modeling language in the automotive domain~\cite{Matinnejad:2016:ATS:2884781.2884797,zander:12}. Simulink appeals to engineers since it is particularly suitable for specifying dynamic systems, and further, it is executable and allows engineers to test their models as early as possible.

\assumption{2}
\emph{Functional requirements are described in a signal logic-based language} (\phase{2}).  
We present our  requirements language in Section~\ref{sec:problemDef} and compare it with existing signal logic languages~\cite{bakhirkin2018first,maler2004monitoring}. We evaluate expressiveness  of our language  in Section~\ref{sec:evaluation}.

\assumption{3}
\emph{A set of test inputs exercising requirements are provided} (\phase{3}).
We assume engineers have a set of test inputs for their CPS model. The test inputs may be generated manually, randomly or based on any test generation framework proposed in the literature~\cite{Matinnejad:2016:ATS:2884781.2884797,zander:12}. Our approach is agnostic to the selected test generation method.

SOCRaTeS automatically converts functional requirements into oracles specified in Simulink (\phase{4}). The oracles evaluate test outputs of the CPS model in an automated and online manner and generate fitness values that provide engineers with a degree of satisfaction or failure for each test input (\phase{5}). Engineers can stop running a test in the middle when SOCRaTeS concludes that the test fitness is going to  remain below a given threshold for the rest of its execution.

\section{Context Formalization}
\label{sec:problemDef}
In Section~\ref{sec:simulinkModel},  we describe CPS Simulink models (\phase{1}) without and with uncertainty and their inputs (\phase{3}). In Section~\ref{sec:FOLfuzzy},  we present \emph{Restricted Signals First-Order Logic (\ourlogic)}, the logic we propose to specify CPS functional requirements (\phase{2}).
In Section~\ref{sec:oracle},   we describe how oracles compute fitness values of test inputs (\phase{5}).

\subsection{Simulink Models}
\label{sec:simulinkModel}
Simulink is a data-flow-based visual language that can be executed using Matlab and consists of blocks, ports and connections.  Blocks typically represent operations and constants, 
and are tagged with ports that specify how data flow in and out of the blocks.
Connections establish data-flows between ports.

\begin{figure}[t]
\includegraphics[width=0.8\columnwidth]{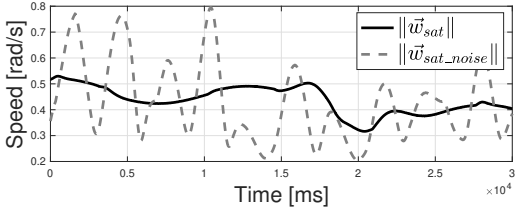}
\caption{Signals $\mathbf{\|} \vec{\mathit{w}}_{\mathit{sat}} \mathbf{\|}$ for the $\mathit{w}_{\mathit{sat}}$  output of \casestudy. The solid-line signal is generated by
 \casestudy\ with no uncertainty, and the dashed-line signal is generated when the S2N ratios in Table~\ref{tab:uncertainParameters} are applied to the \casestudy\ inputs.} 
\label{fig:outputsignal}
\end{figure}

To simulate a Simulink model $M$, the simulation engine receives signal inputs defined over a time domain and computes signal outputs at successive time steps over the same time domain used for the inputs. 
A \emph{time domain} $\timedomain=[0,b]$ is a non-singular bounded interval of \real .  
A \emph{signal} is a function $f: \timedomain \rightarrow \real$. A \emph{simulation}, denoted by $H(\inputs,M)=\outputs$, receives a  set $\inputs=\{\inputsignal_1, \inputsignal_2 \ldots \inputsignal_m\}$ of input signals and produces a set $\outputs=\{\outputsignal_1, \outputsignal_2 \ldots \outputsignal_n\}$ of output signals such that each $o_i \in \outputs$ corresponds to one model output. 
For example, Figure~\ref{fig:outputsignal} shows a signal (black solid line) for the $\mathit{w}_{\mathit{sat}}$ output of     \casestudy\ computed over the time domain $[0, 3\times 10^4]$.

Simulink uses numerical algorithms~\cite{SimulinkSolvers} referred to as \emph{solvers} to  compute simulations.  There are two main types of solvers: fixed-step and variable-step. Fixed-step solvers generate signals over discretized time domains with equal-size time-steps, whereas variable-step solvers  (e.g., Euler, Runge-Kutta~\cite{atkinson2008introduction})  generate signals over continuous time domains.  While the underlying techniques and details of numerical solvers are outside the scope of this paper, we note that our oracles rely on Simulink  solvers  to properly handle signals based on their time domains, whether discrete or continuous.  As a result, our work,  in contrast to existing techniques, is able to seamlessly handle the verification of logical properties over not just discrete but also continuous CPS models.

Simulink has built-in support to  specify and simulate some forms of uncertainty. 
We refer to Simulink models  that contain uncertain elements as \emph{partial} models (denoted $M_p$),
while we use  the term \emph{definitive} to   indicate  models with no uncertainty.
Simulink can specifically capture the following two kinds of uncertainty that are common for CPS and also discussed in Section~\ref{sec:intro} (objective \goal{4})~\cite{uncertainty}:

\emph{(i) Uncertainty due to the noise in inputs.} In Simulink, uncertainty due to the noise is implemented by augmenting model inputs with continuous-time random signals known as \emph{White Noise (WN)}~\cite{golnaraghi2010automatic}. The degree of WN for each input is controlled by a \emph{signal-to-noise ratio (S2N)} value which is the ratio of a desired signal over the background WN~\cite{SignalToNoise}.  Table~\ref{tab:uncertainParameters} shows  the S2N ratios for two inputs of \casestudy.  Fig.~\ref{fig:outputsignal} shows the signal $ \|\vec{\mathit{w}}_{\mathit{sat}}(t)\|$ (gray dashed line) after adding some noise to the original $\mathit{w}_{\mathit{sat}}$ output signal (in black solid line).

\emph{(ii)~Uncertainty related to parameters with unknown values.} In Simulink, parameters whose values are uncertain are typically defined using variables of type 
\emph{uncertain real} (ureal), which is a built-in type in Matlab/Simulink that specifies a range of values for a variable~\cite{ureal}.  Table~\ref{tab:uncertainParameters} shows two parameters of \casestudy\ whose exact values are unknown, and hence, value ranges are assigned to them. 

\begin{figure}
\includegraphics[width=.8\columnwidth]{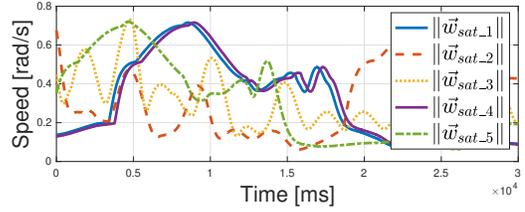}
\caption{A set of signals $\mathbf{\|} \vec{\mathit{w}}_{\mathit{sat}}  \mathbf{\|}$ for the   output $\mathit{w}_{\mathit{sat}}$ of \casestudy\ with uncertain parameters (i.e., when the sun sensor and magnetometer parameters are specified as in Table~\ref{tab:uncertainParameters}).}
\label{fig:simulationPartial}
\end{figure}

Let  $M_p$ be a partial Simulink model with $n$ outputs,  and  let $k$ be the number of different value assignments to  uncertain parameters of $M_p$.  A \emph{simulation}  of a partial Simulink model $M_p$, denoted by  $H_p(\inputs,M_p)=\{\outputs_1,\outputs_2 \ldots \outputs_k\}$, receives a  set $\inputs=\{\inputsignal_1, \inputsignal_2 \ldots \inputsignal_m\}$ of input signals defined over the same time domain, and produces a set of simulation outputs $\{\outputs_1,\outputs_2 \ldots \outputs_k\}$ such  that each $\outputs_i$ is generated by one value assignment to uncertain parameters of $M_p$. Specifically, for each  $O_i \in \{\outputs_1,\outputs_2 \ldots \outputs_k\}$, we have $O_i = \{\outputsignal_1, \outputsignal_2 \ldots \outputsignal_n\}$ such that $o_1, \ldots o_n$ are signals for outputs of $M_p$, i.e., each $O_i$ contains a signal for each output of $M_p$.  The function $H_p$ generates  the simulation outputs consecutively and  is provided in  the Robust Control Toolbox of Simulink~\cite{RobustControlToolbox} which is the uncertainty modeling and simulation tool of Simulink models with  dynamic behavior. 
The value of $k$ indicating the number of value assignments to uncertain parameters can either  be specified by the user or selected based on the recommended  settings of $H_p$.  For example, Figure~\ref{fig:simulationPartial} plots five simulation outputs for the output $\mathit{w}_{\mathit{sat}}$ of \casestudy. The uncertainty in this figure is due to the sun sensor accuracy parameter that  takes values form the range $2.9 \cdot 10^{-3} \pm 10\%$ as indicated in Table~\ref{tab:uncertainParameters}.

\subsection{Our Requirements Language}
\label{sec:FOLfuzzy}

Our choice of a language for CPS requirements is mainly driven by the objectives \goal{1} and \goal{2}   described in Section~\ref{sec:intro}.  These two objectives, however,  are  in conflict. According to \goal{2}, the language should capture complex properties involving magnitude- and time-continuous signals. Such language is expected to have a high runtime computational complexity~\cite{Bakhirkin:2018:FLS:3283535.3283536}. This, however, makes the language unsuitable for the  description of online oracles that should typically have low runtime computational complexity, thus contradicting \goal{1}. For example, Signals First Order (SFO) logic~\cite{Bakhirkin:2018:FLS:3283535.3283536} is an extension of  first order logic with continuous signal variables. SFO, however, is not amenable to online checking in its entirety due to its high expressive power that leads to high computational complexity of monitoring SFO properties~\cite{Bakhirkin:2018:FLS:3283535.3283536}.  Thus, the procedure for monitoring SFO properties is  tailored to offline checking.  In order to achieve both \goal{1} and \goal{2}, we  define Restricted Signals First-Order Logic (\ourlogic), a fragment of SFO. \ourlogic\ can be effectively mapped to Simulink to generate online oracles that run together with the model under test  by the same solvers applied to the model, which can handle any signal type (i.e., discrete or continuous), hence addressing both \goal{1} and \goal{2}. Note that even though \ourlogic\ is less expressive than SFO, as we will discuss in Section~\ref{sec:evaluation}, all CPS requirements in our case studies can be captured by  \ourlogic.

\emph{\ourlogic\  Syntax.} 
Let $T=\{ t_1, t_2, \ldots t_d\}$ be a set of \emph{time variables}.
 Let $F = \{f_1, f_2, \ldots, f_l\}$ be a set of signals defined over the same time domain \timedomain, i.e., $f_i: \timedomain \rightarrow \real$ for every $1 \leq i \leq l$.  

Let us consider the grammar $\mathcal{G}$ defined as follows:

\scalebox{.95}{\parbox{.5\linewidth}{
\begin{align}
&\tau  \Coloneqq && t + n \mid  t - n  \mid t \mid n &  \nonumber \\
&\rho   \Coloneqq &&   f(\tau) \mid \textbf{g}(\rho) \mid \textbf{h}(\rho_1,\rho_2) & \nonumber \\
&\phi  \Coloneqq && \rho  \sim r   \mid \phi_1 \vee \phi_2 \mid \phi_1 \wedge \phi_2 \mid
\forall t \in \langle \tau_1, \tau_2 \rangle \colon \phi   \mid
\exists t \in \langle \tau_1, \tau_2 \rangle \colon \phi & \nonumber 
\end{align}}}

where $n \in \real_0^+ $, $t \in T$, $f \in F$, $r \in \real$, and $\textbf{g}$ and $\textbf{h}$ are, respectively,  arbitrary unary and binary arithmetic operators, $\sim$ is a relational operator in $\{< , \leq, >, \geq, =, \neq \}$, and $\langle \tau_1, \tau_2 \rangle$ is a \emph{time interval} of \timedomain\ (i.e., $\langle \tau_1, \tau_2 \rangle \subseteq \timedomain$) with lower bound $\tau_1$ and upper bound $\tau_2$. 
The symbols $\langle$ and $\rangle$  are equal to $[$ or $($, respectively to $]$ or $)$, depending on whether  $\tau_1$, respectively  $\tau_2$, are included  or excluded from the interval. 
We refer to  $\tau$, $\rho$ and $\phi$  as \emph{time term}, \emph{signal term} and \emph{formula term}, respectively.
A \emph{predicate} is a formula term in the form $\rho  \sim r$.

\begin{definition}\label{def:rfol}
A  \emph{Restricted Signals First-Order Logic}  (\ourlogic) formula  $\varphi$  is a formula term defined according to the grammar $\mathcal{G}$ that also satisfies the following conditions: (1)~$\varphi$ is closed,  i.e., it does not have any free variable; and (2)~every sub-formula of $\varphi$ has at most one free time variable. 
\end{definition}

In \ourlogic, boolean operations ($\wedge$, $\vee$) combine predicates of the form $ \rho  \sim r$,  which compare signal terms with real values\footnote{Note that in our logic, negation $\neg$ is applied at the level of predicates.}. 
The formulas further quantify over time variables of signal terms $\rho$ in $\rho \sim r$ and bound them in time intervals $\langle \tau_1, \tau_2 \rangle$. Table~\ref{tab:requirements} shows the formalization of the  \casestudy{} requirements  in \ourlogic{}.  For example,  the predicate  $\|\vec{\mathit{w}}_{\mathit{sat}} \|<1.5 $ of 
formula $R1$ states that the angular velocity of the satellite should  be less than 1.5m/s,
and  $\forall t \in [0, \numprint{86400})$ forces the predicate to hold for a duration of \numprint{86400}s $\simeq 24$h,  the estimated time required for the satellite to finish an orbit.

\emph{\ourlogic\ expressiveness.} Here, we discuss what types of SFO properties 
are eliminated from   \ourlogic\  due to the conditions in Definition~\ref{def:rfol}. 
Condition~1 in Definition~\ref{def:rfol} requires closed formulas.  \ourlogic\ properties must not  include free variables (i.e., they should be formulas and not queries) so that they generate definitive results when checking test outputs. Condition~2 in Definition~\ref{def:rfol}  is needed to  ensure that the formulas can be translated into online oracles specified in Simulink.  This condition eliminates formulas containing predicates $\rho \sim r$ where $\rho$ includes an arithmetic operator applied to signal segments over different time intervals (i.e., signal segments with different time scopes). For example, the formula $\forall t \in [1,5] : \forall t' \in [7,9] : f(t) + f(t') < 4$ is not in \ourlogic\ since $f(t) + f(t') < 4$ has two free time variables $t$ and $t'$ (i.e., it violates condition 2 in  Definition~\ref{def:rfol}). The predicate $f(t) + f(t') < 4$ in this formula computes the sum of two segments of signal $f$ related to time intervals $[1,5]$ and $[7,9]$. Such formulas are excluded from \ourlogic\   since during online checking, the operands $f(t)$ and
$f(t')$ cannot be simultaneously accessed to compute $f(t) + f(t')$.  We note that formulas with arithmetic operators applied to signal segments over the \emph{same} time interval (e.g., R4 and R5 in  Table~\ref{tab:requirements}), or formulas involving different predicates over different time intervals, but connected with \emph{logical} operators (e.g., R6) are included in \ourlogic.

\emph{Comparison with STL.}  In addition to SFO, Signal Temporal Logic (STL)~\cite{maler2004monitoring} is another logic  proposed  in the literature that can capture CPS continuous behaviors. We compare \ourlogic\ with STL, and in particular, with \emph{bounded} STL  since test oracles can only check signals generated up to a given bound. Hence, for our purpose, bounded STL  temporal operators have to be applied (e.g., $\LTLu_{[a,b]}$).  \ourlogic\  subsumes bounded STL since boolean operators of STL can be trivially expressed in \ourlogic, and any temporal STL formula  in the form of $\varphi_1 \LTLu_{[a,b]} \varphi_2 $ can also be specified in \ourlogic\ using time terms and time intervals.  The detailed translation is available online~\cite{AdditionalMaterial}.

\emph{\ourlogic\  Semantics.}  We propose a  (quantitative) semantics for \ourlogic\ to help engineers distinguish between different degrees of satisfaction and failure (see objective \goal{3} in Section~\ref{sec:intro}).   As shown in Table~\ref{tab:requirements} and also based on \ourlogic\  syntax, CPS requirements essentially  check predicates  $\rho \sim r$  over time. In other words, predicates  $\rho \sim r$  are the building blocks of \ourlogic. To define a quantitative semantics for \ourlogic, we need to first define the semantics of these predicts in a quantitative way. In our work, we define a (domain-specific) \diff\ function to assign a fitness value to $\rho \sim r$. We require \diff\ to have these characteristics:  (1)~The range of \diff\ is $[-1,1]$. (2)~A value in $[0,1]$ indicates that  $\rho \sim r$ holds,  and a value in $[-1,0)$ indicates that $\rho \sim r$ is violated.

\begin{definition}
\label{def:semantics}
Let $\diff$\ be a domain-specific semantics function for predicates $\rho \sim r$. 
Let $F  = \{f_1, \ldots, f_l\}$ be a set of signals with the same time domain \timedomain.   The semantics of an  \ourlogic\ formula $\phi$ for the signal set $F$  is denoted by $\interpretation{\phi}_{F}$ and  is  defined as follows: 
\scalebox{.9}{\parbox{.5\linewidth}{
\begin{align*}
& \interpretation{f(n)}_{F} && = && \begin{cases}
 f(n) &  \text{  if  $f \in F$  and  $n \in \timedomain$} \\
\text{undefined} &  \text{ otherwise}
\end{cases} 
 & & \nonumber \\
 &  \interpretation{\textbf{g}(\rho)}_{F}&& = &&   \textbf{g}(\interpretation{\rho}_{F}) && \nonumber \\
&  \interpretation{\textbf{h}(\rho_1,  \rho_2)}_{F} && = &&  \textbf{h}(\interpretation{\rho_1}_{F},  \interpretation{\rho_2}_{F}) && \nonumber \\
& \interpretation{\rho \sim r}_{F} && = && \diff(\interpretation{\rho}_{F} \sim r) && \nonumber \\
&  \interpretation{\phi_1 \wedge \phi_2}_{F}  && =&&  min( \interpretation{\phi_1}_{F}, \interpretation{\phi_2}_{F}) && \nonumber \\
&  \interpretation{\phi_1 \vee \phi_2}_{F}  && =&& max( \interpretation{\phi_1}_{F}, \interpretation{\phi_2}_{F})
&& \nonumber \\
&\interpretation{\forall t \in \langle n_1, n_2 \rangle \colon \phi}_{F} &&=&&   
\underset{\forall t' \in \langle n_1, n_2  \rangle}{min}(\interpretation{\phi[t \leftarrow t']}_{F} ) 
&& \nonumber \\
&\interpretation{\exists t \in \langle n_1, n_2 \rangle \colon \phi}_{F} &&=&&   
\underset{\forall t' \in \langle n_1, n_2  \rangle}{max}(\interpretation{\phi[t \leftarrow t']}_{F} ) 
&& \nonumber 
\end{align*}}}
\end{definition}

The choice of the $max$ and $min$ operators for defining the semantics of $\exists$ and $\forall$ 
is standard~\cite{larsen1988modal}: 
the minimum has the same  behavior as  $\wedge$  and evaluates  whether a predicate holds over the entire time  interval. Dually,  the max operator captures $\vee$.  The semantics of signal terms $f(n)$ depends on whether the signal is included in $F$ and whether $n$ is in the time domain  \timedomain, otherwise $f(n)$ is undefined. We say $\varphi \in \ourlogic$ is \emph{well-defined with respect to a signal set $F$} iff  no signal term in $\varphi$ is undefined.  To avoid undefined \ourlogic\ formulas, signal time domains $\timedomain$ should be selected such that signal indices are included in \timedomain, and further, the formula should not have negative signal indices. For example, for properties in Table~\ref{tab:requirements}, we need a time domain  $\timedomain = [0,\numprint{86400}]$ for {\bf R1} to {\bf R4}, a time domain  $\timedomain = [0,\numprint{86402}]$ for {\bf R5}, and a time domain $\timedomain = [0,\numprint{88400}]$ for {\bf R6}.  Finally, we can infer the boolean semantics of \ourlogic\ from its quantitative semantics:  For every formula term $\varphi$, we have $F \models \varphi \mbox{ iff } \interpretation{\varphi}_{F} \geq0$. In other words, $\varphi$ holds over the signal set $F$ iff $\interpretation{\varphi}_{F} \geq 0$.

Let $\mu =  \interpretation{\rho}_F - r$.  In our work, we define \diff\  as follows: 

\scalebox{.9}{\parbox{.5\linewidth}{
\begin{align*}
&\diff(\interpretation{\rho}_F = r )   =  \frac{- |\mu|}{|\mu| +1} 
&&  \diff(\interpretation{\rho}_F \neq r )   =  \begin{cases} \frac{|\mu|}{|\mu| +1}  &  \text{ if $\mu \neq 0$} \\
-\epsilon &  \text{else} \end{cases} &  \nonumber \\
&\diff(\interpretation{\rho}_F \geq r )   =   \frac{\mu}{|\mu|+1}  
&& \diff(\interpretation{\rho}_F > r ) =   \begin{cases} \frac{\mu}{|\mu| +1}  &  
\text{ if $\mu \neq 0$} \\
-\epsilon &  \text{else} \end{cases} &  \nonumber \\
&\diff(\interpretation{\rho}_F \leq r )  =   \frac{- \mu}{|\mu|+1}  
&& \diff(\interpretation{\rho}_F < r )  =   \begin{cases} \frac{- \mu}{|\mu|+1}  &  
\text{ if $\mu \neq 0$} \\
-\epsilon &  \text{else} \end{cases}
 &  \nonumber 
\end{align*}}}

In the above,  $\epsilon$ is an infinitesimal positive value that ensures $\diff <0$ when $\mu = 0$ and either $<$, $>$ or $\neq$ is used.

Our \diff\ function satisfies the two conditions described earlier and is closed under logical $\wedge$ and $\vee$. For example, $(\rho \leq r) \wedge (\rho \geq r)$ is equal to $(\rho = r)$.  Our \diff\ function, further, provides a quantitative fitness measure distinguishing  between different levels of satisfaction and refutation. Specifically, a  higher value of \diff\ indicates that $\rho \sim r$ is fitter (i.e., it better satisfies or less severely violates the requirement under analysis).  For example, the  \diff\  value of the predicate  $\|\vec{\mathit{w}}_{\mathit{sat}} \|<1.5$ for the signals shown in Figure~\ref{fig:simulationPartial} is above zero implying that the signals satisfy the predicate. In contrast, the \diff\ values for signals $\|\vec{q}_{real}-\vec{q}_{target}\|$ in Figures~\ref{fig:motivation}(b) and (c) are $-0.5$ and $-0.95$, respectively. This shows that the violation in Figure~\ref{fig:motivation}(c) is more severe than that in Figure~\ref{fig:motivation}(b). 

Note that  the above $\diff$\  function is only one alternative where we assume the fitness is proportional to the difference between $\rho$ and $r$. We can define  the \diff\ function differently as long as the two properties described earlier are respected and the proposed semantics for \diff\ respects logical conjunction and disjunction operators. For example, a simple \diff\  can return $1$ when $\rho \sim r$ holds and -1, otherwise, yielding the boolean semantics of \ourlogic.

\subsection{Test Oracles}
\label{sec:oracle}
In this section, we formally define our notion of test oracle. We specifically discuss test oracles for partial Simulink models since a definitive model is a specialization of a partial model. Recall that by simulating a partial Simulink model $M_p$ for a given test input $I$, we obtain a set of $k$ alternative signals for each output of $M_p$, while  for a definitive model $M$, the simulation output contains only one signal for each model output.

\begin{definition}
\label{def:oracle}
Let $M_p$ be a Simulink model under test, and let $I$ be a test input for $M_p$ defined over the time domain $\timedomain$. Let $\varphi$ be an \ourlogic\ formula formalizing a requirement of $M_p$. Suppose $\{\outputs_1,\outputs_2 \ldots \outputs_k\}=H_p(\inputs,M_p)$ are the simulation results generated for the time domain $\timedomain$.  We denote the \emph{oracle} value of $\varphi$  for test input $I$ over model $M_p$  by  \hbox{$\text{oracle}(M_p, I , \varphi)$} and compute it as follows:
\begin{align}
\text{oracle}(M_p, I , \varphi) =\underset{O \in \{\outputs_1,\outputs_2 \ldots \outputs_k\}}{min}\interpretation{\varphi}_{O} \nonumber
\end{align} 
Specifically, $\text{oracle}(M_p, I , \varphi)$ indicates the fitness value of the test input $I$  over model $M_p$ and evaluated against  requirement $\varphi$.  
\end{definition}

Recall that based on Definition~\ref{def:semantics}, the oracle output is a value in $[-1,1]$. For definitive models,  the test yields a single set $O = \{o_1, \ldots, o_n\}$ of simulation outputs, and hence,  the oracle computes $\interpretation{\varphi}_{O}$, i.e., it evaluates $\varphi$ over the set $O$ of test outputs. As defined above, for a partial model, the oracle computes the minimum value of $\varphi$ over every test output set. Hence, for a partial model, the fitness value for a test $I$ is determined by the model output yielding the lowest fitness  (i.e., the model output revealing the most severe failure or the model output yielding the lowest passable fitness).

\section{Oracle generation}
\label{sec:contribution}
In this section, we present the oracle generation component of  SOCRaTes (\phase{4} in Figure~\ref{fig:approach}).  This component automatically translates \ourlogic\ formulas into \emph{online} test oracles specified in Simulink that can handle time and magnitude-continuous signals and conform to our notion of oracle described in Definition~\ref{def:oracle}.  Note that an \ourlogic\  formula may not be \emph{directly} translatable into an online test oracle if it contains sub-formulas referring to future time instants or to signal values that are not yet generated at the current simulation time. 
For example,  consider the predicate $ \| \vec{q}_{real}(t+2000)-\vec{q}_{estimate}(t+2000) \| \leq 0.02$ in the  \textbf{R6} property of Table~\ref{tab:requirements}. The fitness value of this predicate at $t$ (i.e., the oracle output in Definition~\ref{def:oracle}) can only be  evaluated after generating signals  $\vec{q}_{real}$ and $\vec{q}_{estimate}$  up to  the time instant $t+2000$. This requires extending the time domain \timedomain\ by $2000$ seconds. Instead of forcing a longer simulation time, we propose a procedure that rewrites the \ourlogic\ formulas into a form that allows a direct translation into online test oracles.
This procedure, called \emph{time and interval shifting},  is presented in Section~\ref{sec:timeshifting}. Having applied the procedure to  \ourlogic\ formulas, in Section~\ref{sec:SFFO2simulink}, we describe our translation to convert \ourlogic\ formulas into  \emph{Simulink oracles}. We further present a  proof of soundness and completeness of our translation  in that section. All the proofs of the Theorems are provided in our online Appendix~\cite{AdditionalMaterial}.

\subsection{Time and Interval Shifting}
\label{sec:timeshifting}

Below, we present the \emph{time-} and  \emph{interval-shifting} steps separately:

\emph{Time-shifting.} Any signal term that refers to a signal value generated in the future should be rewritten as a signal term that  does not refer to the future. For example, the formula $\vec{q}_{real}(t+2000)<5$ that refers to the value of  $\vec{q}_{real}$ in  the future cannot be checked online. Therefore, our time-shifting procedure  replaces any signal term $f(t+n)$  with a signal term  $f(t-n)$ as follows:  Let $\psi$ be an \ourlogic\ formula.  We traverse $\psi$ from its leaves to its root and replace every sub-formula  \hbox{$\forall t \in \langle n_1,n_2 \rangle \colon \phi(t)$} (resp. $\exists t \in \langle n_1,n_2 \rangle \colon \phi(t)$)  of $\psi$  with  \hbox{$\forall t \in \langle n_1 + d_t,n_2 +d_t \rangle \colon \phi(t - d_t)$} (resp. $\exists t \in \langle n_1 + d_t,n_2 +d_t \rangle \colon \phi(t - d_t)$),  where $d_t$ is the maximum value of constant $n$ in time terms $t+n$ appearing as signal indices in $\phi(t)$.  For example, the requirement {\bf R5} in Table~\ref{tab:requirements} is rewritten as:
\newline \hbox{$\forall t \in [2, \numprint{86 402})  \colon  \| \vec{q}_{target}(t-2)-\vec{q}_{target}(t) \| \leq 2\times sin(\frac{\alpha}{2})$}

\emph{Interval-Shifting.}  To ensure that $\psi$ can be translated into an online test oracle, for  any  $\forall t \in \langle \tau_1, \tau_2 \rangle: \phi$ in $\psi$, the interval $\langle \tau_1, \tau_2 \rangle$ should end after all the intervals 
$\langle\tau'_1, \tau'_2 \rangle$ such that $\forall t' \in \langle\tau'_1, \tau'_2 \rangle:\phi'$ is a sub-formula of $\phi$ (i.e., $\tau_2 \geq \tau'_2$),  and further, it should begin after all the intervals 
$\langle\tau'_1, \tau'_2 \rangle$ such that $\exists t'\in \langle\tau'_1, \tau'_2 \rangle:\phi'$ is a sub-formula of $\phi$ (i.e., $\tau_1 \geq \tau'_2$). Similarly,  for  any  $\exists t \in \langle \tau_1, \tau_2 \rangle: \phi$ in $\psi$, the dual of the above two conditions must hold. These conditions will ensure that the evaluation of the sub-formulas in the scope of $t$ can be fully contained and completed within the evaluation of their outer formula.  For example, $\forall t \in [0,3]   \colon f(t) = 0 \wedge \forall t' \in [0,5]   \colon f(t') = 1$ cannot be checked in an online way since the time interval  of the inner sub-formula (i.e., $[0,5]$) does not end before the time interval  of the outer formula (i.e., $[0,3]$). Therefore, our interval-shifting procedure shifts each time interval $\langle \tau_1, \tau_2\rangle$ to ensure that it terminates after all its related inner time intervals. 

Let  $\psi$  be an \ourlogic\ formula. We traverse $\psi$ from its leaves to its root and we perform the following operations: 
(i) replace every sub-formula
 $\forall t \in \langle\tau_1,\tau_2\rangle\colon \phi(t)$  of $\psi$ with $\forall t \in \langle\tau_1 + d_u,\tau_2 +d_u\rangle \colon \phi(t-d_u)$,  where $d_u$ is the maximum value of constant $n$ in   the upper bounds $\tau_2$ of  time intervals $\langle\tau_1, \tau_2\rangle$ associated with $\forall$ operators
and the lower bounds $\tau_1$ of time intervals $\langle\tau_1, \tau_2\rangle$ associated with $\exists$ operators
in $\phi(t)$; 
(ii) execute a dual procedure to update the time intervals of existential sub-formulae.
For example, the interval-shifting procedure rewrites the formula previously introduced as 
 $\forall t \in [2,5]   \colon f(t-2) = 0 \wedge \forall t' \in [0,5]   \colon f(t') = 1$.

To ensure interval-shifting is applied to signal variables with constant indices, we replace every $f(n)$  in $\psi$ where $n$ is a constant  with $\forall t^* \in [n,n]: f(t^*)$ where $t^*$ is a new time variable that has not been used in $\psi$.  We refer to the  \ourlogic\  formula  obtained by sequentially applying time-shifting and  interval-shifting  to an \ourlogic\ formula $\varphi$ as \emph{shifted-formula}  and denote it by $\formula_\Uparrow$.

\begin{restatable}{theorem}{theoremshifting}
\label{thm:shifting}
Let \formula\ be an \ourlogic\ formula and let $\formula_\Uparrow$ be its \emph{shifted-formula}. 
 For any signal set $F$, we have:  $\interpretation{\formula}_{F} = \interpretation{\formula_{\Uparrow}}_{F}$

\end{restatable}

The time complexity of generating a \emph{shifted-formula}  $\formula_\Uparrow$ is $ |\formula|$ where $|\formula|$ is the size of the formula   \formula,  i.e., the sum of the number of its temporal and arithmetic operators. Both time and the interval shiftings scan the syntax tree of $\varphi$ from  its leaves to the root twice: one for computing the shifting values $d_t$ and $d_u$ for every subformula of $\varphi$; and the other to apply the shifting, i.e., replacing the variable $t$ with $t-d_t$ or $t-d_u$.

\subsection{From \ourlogic\ to Simulink}
\label{sec:SFFO2simulink}
In this section, we translate \ourlogic\ formulas written in their shifted-forms (as described in Section~\ref{sec:timeshifting})  into Simulink.   Table~\ref{tab:simulinkOracleGeneration} presents the rules for translating each syntactic construct of \ourlogic\ defined in Definition~\ref{def:rfol} into Simulink blocks. Note that \textbf{h} and \textbf{g} in Table~\ref{tab:simulinkOracleGeneration}, respectively, refer to binary arithmetic operators (e.g., +) or unary functions (e.g., \texttt{sin}) and map to their corresponding Simulink operations. Below, we discuss the rules for  $t$, $f(t-n)$,   
$\forall t \in \langle \tau_1, \tau_2 \rangle \colon \phi$, and $\rho \sim r$ since the other rules in Table~\ref{tab:simulinkOracleGeneration} directly  follow from the RFOL semantics. Note that  signal variables in shifted formulas  are all written as $f(t -n)$ s.t. $n \geq 0$.  Hence, we give a translation rule for signal variables in the form of $f(t-n)$ only.

\begin{table*}[h]
\caption{Translating the SFFO formulae into Simulink Oracles.}
\label{tab:simulinkOracleGeneration}
\scalebox{.9}{\begin{tabular}{| p{.07\textwidth}  | p{.08\textwidth} | p{.08\textwidth} | p{.12\textwidth} |p{.2\textwidth} | p{.1\textwidth} |}
\toprule
\textbf{Rule} & 
\multicolumn{1}{c |}{Rule1} &
\multicolumn{1}{c |}{Rule2} &
\multicolumn{1}{c |}{Rule3} &
\multicolumn{1}{c |}{Rule4} &
\multicolumn{1}{c |}{Rule5}
\\  
\midrule
\textbf{Formula} & 
\multicolumn{1}{c |}{$t$} &
\multicolumn{1}{c |}{$n$} &
\multicolumn{1}{c |}{$t\pm n$} &
\multicolumn{1}{c |}{$f(t - n)$} &
\multicolumn{1}{c |}{$\textbf{h}(\rho_1, \rho_2)$/$\textbf{g}(\rho)$}  
\\  
\midrule
\textbf{Simulink} & 
\multicolumn{1}{c|}{\raisebox{-0.75\totalheight}{
\includegraphics[scale=0.45]{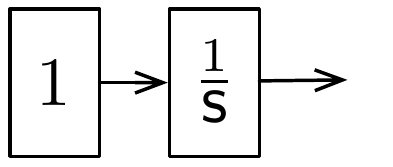}}}
&
\multicolumn{1}{c|}{\raisebox{-0.75\totalheight}{
\includegraphics[scale=0.45]{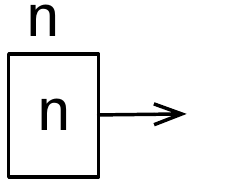}}} 
&
\multicolumn{1}{c|}{\raisebox{-0.75\totalheight}{
\includegraphics[scale=0.45]{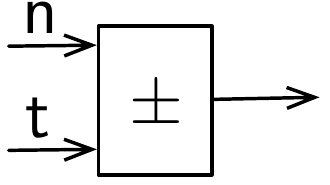}}}
&
\multicolumn{1}{c|}{\raisebox{-0.75\totalheight}{
\includegraphics[scale=0.45]{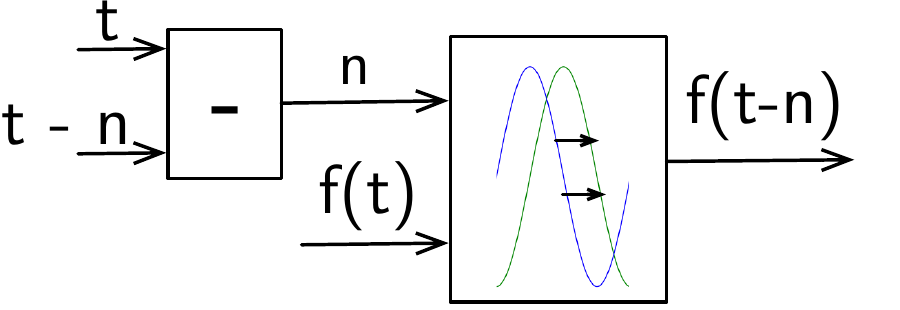}}}  
&
\multicolumn{1}{c|}{\raisebox{-0.75\totalheight}{\includegraphics[scale=0.5]{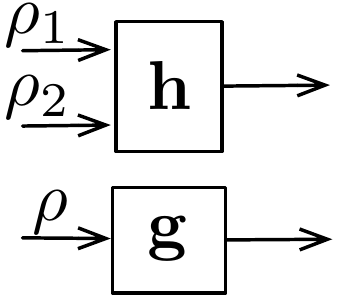}}}
 \\ 
 \toprule
 \textbf{Rule} & 
\multicolumn{1}{c |}{Rule6}  &
\multicolumn{1}{c |}{Rule7}  &
\multicolumn{2}{c |}{Rule8} &
\multicolumn{1}{c |}{Rule9}
\\  
\midrule
\textbf{Formula} & 
\multicolumn{1}{c |}{$\phi_1 \vee \phi_2$} & 
\multicolumn{1}{c |}{$\phi_1 \wedge \phi_2$}& 
\multicolumn{2}{c |}{$\forall t \in \langle\tau_1, \tau_2\rangle \colon  \phi$}& 
\multicolumn{1}{c |}{$\rho \sim r$}\\
\midrule  
\textbf{Simulink} &  
\multicolumn{1}{c|}{\raisebox{-0.75\totalheight}{
\includegraphics[scale=0.5]{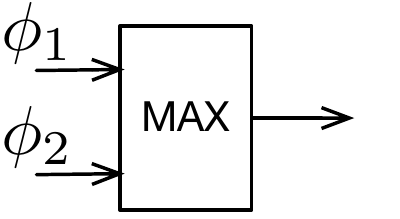}} } &  
\multicolumn{1}{c|}{\raisebox{-0.75\totalheight}{
\includegraphics[scale=0.5]{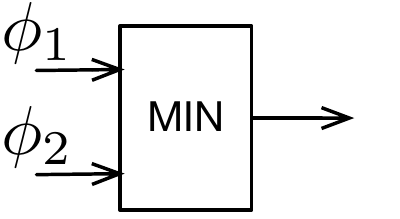}} } & 
\multicolumn{2}{c|}{\raisebox{-0.75\totalheight}{
\includegraphics[scale=0.5]{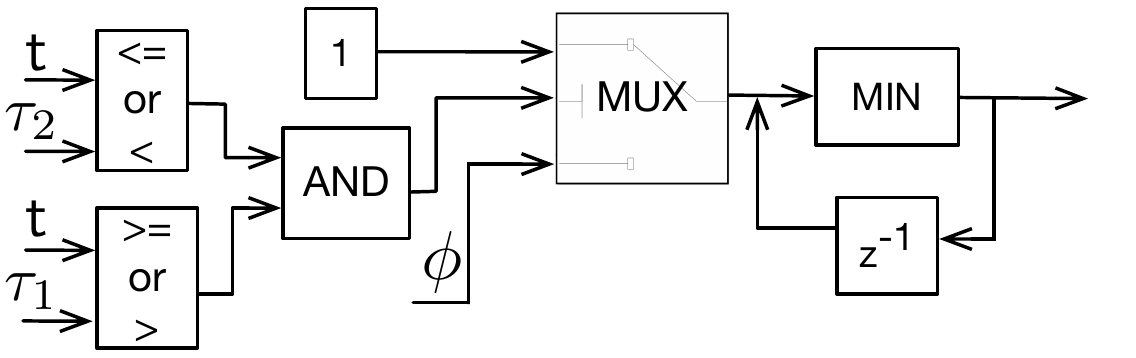}}}  &
\multicolumn{1}{c|}{\raisebox{-0.75\totalheight}{
\includegraphics[scale=0.5]{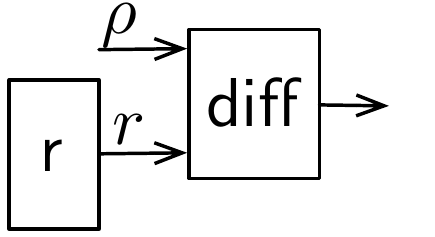}} }
\\ 
 \bottomrule
\end{tabular}}
\end{table*}

\noindent $\bullet$ Rule1: To compute the value of $t$, we use an integrator Simulink block to compute  the formula 
$\int_{0}^{t}  dt$ which yields $t$.  \\
$\bullet$ Rule4:  To encode $f(t - n)$, we first obtain the delay $n$ applied to the signal $f$. To obtain the value of $n$ from $t-n$,  we compute  $t - (t - n)$. We then use the \emph{transport delay} block of Simulink to obtain the value of $f$ at $n$ time instants before $t$. \\
$\bullet$ Rule8: The formula  $\forall t \in \langle \tau_1, \tau_2 \rangle \colon \phi$ is mapped into a Simulink model that initially generates the value $1$ until the start of the time interval $\langle \tau_1, \tau_2 \rangle$. 
When $t \in \langle \tau_1, \tau_2 \rangle$ holds, the multiplexer of Rule8 selects the value of $\phi$ instead of $1$. 
Note that we use symbol $<$ for $\langle = ``("$, 
symbol $\leq$ for $\langle = ``["$, 
symbol $>$ for $\rangle = ``)"$, 
and symbol $\geq$ for $\rangle = ``]"$. 
The feedback loop in the model  combined with a delay block (i.e., z$^{-1}$) computes the minimum of $\phi$ over the time interval  $\langle \tau_1, \tau_2 \rangle$. 
Once the time interval  $\langle \tau_1, \tau_2 \rangle$ expires, the multiplexer chooses constant 1 again.  This, however, has no side-effect on the value $v$ already computed for the formula $\forall t \in \langle \tau_1, \tau_2 \rangle \colon \phi$ because $v \leq 1$ and the minimum of $v$ and $1$ remains $v$ until the end of the simulation.  Note that the rule for translating $\exists t \in \langle \tau_1, \tau_2 \rangle \colon \phi$ into Simulink is simply obtained by replacing in Rule8 \textsf{MIN} with \textsf{MAX} and constant 1 with  constant -1. \\
$\bullet$ Rule9: Recall that the semantics of $\rho \sim r$ depends on a domain specific fitness function. In our work,   we implement the \textsf{diff} block in Rule9 based on the functions given in \hbox{Section~\ref{sec:FOLfuzzy} for function \diff.}

Let $\varphi$  be an \ourlogic\ formula and $\varphi_\Uparrow$  be its corresponding shifted formula.  We denote by $M_\varphi$  the Simulink model obtained by translating  $\varphi_\Uparrow$ using the rules in Table~\ref{tab:simulinkOracleGeneration}. 
The model $M_\varphi$ is a definitive Simulink model and has one and only one output because every model fragment in  Table~\ref{tab:simulinkOracleGeneration} has one single output.  This output will be indicated in the following with the symbol $e$. 
Below, we argue that $M_\varphi$ conforms to our notion of test oracle given in Definition~\ref{def:oracle}, and  is an online oracle that can handle continuous signals.  In order to use  $M_\varphi$ to check outputs of model $M_p$  with respect to a property $\varphi$, it suffices to connect the outputs of $M_p$ to the inputs of $M_\varphi$. We denote the model obtained by connecting the output ports of $M_p$ to the input ports of $M_\varphi$ by $M_p + M_{\varphi}$. Clearly,  $M_p + M_{\varphi}$ has only one output signal $e$ (i.e., the output of $M_\varphi$).

\begin{restatable}{theorem}{theoremtranslation}
\label{thm:correctness}
Let $M_p$ be a (partial) Simulink model, and let
$I$ be a test input for $M_p$ defined over the time domain $\timedomain = [0,t_u]$. 
Let $\varphi$ be a requirement of $M_p$  in \ourlogic. Suppose $\{\outputs_1,\outputs_2 \ldots \outputs_k\}=H_p(\inputs,M_p)$ and  $\{\{e_1\}, \{e_2\}, \ldots, \{e_k\}\}=H_p(I ,M_p+\testoracle)$ are simulation results generated for the time domain $\timedomain$.   Then, the value of $\varphi$ over every signal set $O_i \in \{\outputs_1,\outputs_2 \ldots \outputs_k\}$ is equal to the value of the signal $e_i$ generated by $M_p+\testoracle$ at time $t_u$. That is, $\interpretation{\varphi}_{O_i} = e_i(t_u)$. Further, we have:
\begin{align}
\text{oracle}(M_p, I, \varphi) =\underset{e \in \{e_1, \ldots, e_k\}}{min} e(t_u) 
\nonumber
\end{align} 
That is, the minimum value of  the outputs of  $M_p+\testoracle$ at $t_u$ is equal to  the oracle value  as defined by Definition~\ref{def:oracle}. 
\end{restatable}

Theorem 4.2  states that our translation of \ourlogic\ formulas  into Simulink is \emph{sound} and \emph{complete} with respect to our notion of oracle in Definition~\ref{def:oracle}. Note that in the case of a definite Simulink model $M$, the output of $M + M_\varphi$ is a single signal $e$. In summary, according to  Theorem 4.2, $M_p + M_\varphi$ (or $M + M_\varphi$)  is able to correctly compute the fitness  value of $\varphi$ for test input $I$.

\begin{restatable}{theorem}{theoremstop}
Let $M_p$ be a (partial) Simulink model, and let
$I$ be a test input for $M_p$ over the time domain $\timedomain=[0, t_u]$. 
Let $\varphi$ be a requirement of $M_p$  in \ourlogic. 
Suppose $\{\{e_1\}, \{e_2\}, \ldots, \{e_k\}\}=H_p(I ,M_p+\testoracle)$ are simulation results generated for  $\timedomain$.   Let $d$ be the maximum constant appearing in the upper bounds of the time intervals of $\varphi_{\Uparrow}$ for existential quantifiers (i.e., time intervals in the form of $\exists t \in  [\tau_1, \tau_2] : \phi$ in $\varphi_{\Uparrow}$).  Each $e_i \in \{\{e_1\}, \{e_2\}, \ldots, \{e_k\}\}$  is decreasing over the time interval  
$(d,t_u]$. 
\end{restatable}

Note that $d$ in Theorem 4.3 indicates the time instant when all the existentially quantified time intervals  of $\varphi$ are terminated, and hence all the sub-formulas within the existential quantifiers of $\varphi$ are evaluated.  According to Theorem 4.3, the oracle output for $\varphi$ becomes monotonically decreasing after $d$. Therefore,  after $d$, we can stop model simulations as soon as the output of  $M_p + M_\varphi$  falls below some desired threshold level. More specifically, if the  output of $M_p + M_\varphi$ falls below a threshold at time $t > d$ it will remain below that threshold for any $t^\prime \geq t$. Hence, $M_p + M_\varphi$  is able to check test outputs in an online manner and stop simulations within the time interval $(d,t_u]$ as soon as some undesired results are detected.  Note that $d = 0$ if $\varphi$ does not have any existential quantifier.

Our oracles can check Simulink models with time and magnitude-continuous signal outputs since  all the blocks used in Table~\ref{tab:simulinkOracleGeneration} can be executed by both fixed-step and variable-step solvers of Simulink, where the time step is decided by the same solver applied to the model under test. Finally, the running time of our oracle is linear in the size of the underlying time domain $\timedomain$.

\section{Evaluation}
\label{sec:evaluation}
In this section, we empirically evaluate SOCRaTEs using  eleven realistic and industrial  Simulink models from the CPS domain. Specifically, we aim to answer the following questions.  \textbf{RQ1:} Is our requirements language (\ourlogic) able to capture CPS requirements in industrial settings? \textbf{RQ2:} Is the use of RFOL and our proposed translation into Simulink models likely to be practical and beneficial?  \textbf{RQ3:} Is a  significant amount of execution time saved when using online test oracles, as compared to offline checking?

\emph{Implementation.} We implemented SOCRaTEs as an Eclipse plugin using Xtext~\cite{Xtext} and Sirius~\cite{sirius} and have made it available online~\cite{SOCRaTEs}.

\emph{Study Subjects.}  We evaluate our approach using eleven case studies listed in  Table~\ref{table:cpsmodels}. We received the case studies from two industry partners: LuxSpace~\cite{Luxspace},  a satellite system developer, and QRA Corp~\cite{QRA}, a verification tool vendor to the aerospace, automotive and defense sectors. Each case study includes a Simulink model  and a set of functional requirements  in natural language that must be satisfied by the model.  Two of our case studies, i.e., \casestudy\ from LuxSpace and \textsf{Autopilot} from QRA Corp, are large-scale industrial models and respectively represent full behaviors  of a satellite and an autopilot system and their environment.  The other nine models capture smaller systems or sub-systems of some CPS.   Our case study models implement diverse CPS functions  and capture complex behaviors such as non-linear and differential equations, continuous behaviors and uncertainty.   \casestudy\ and \textsf{Autopilot} are continuous models. \casestudy\ further has inputs with noise and some parameters with uncertain values.  Table~\ref{table:cpsmodels} also reports the number of blocks (\#Blocks) of the Simulink models and the number of requirements (\#Reqs) in our case studies.  In total, our case studies include 98 requirements.

\begin{table*}[t]
\caption{Important characteristics of our case study systems (from left to right): (1)~name, (2)~description, (3)~number of blocks of the Simulink model of each case study ({\bf \#Blocks}), and (4)~number of requirements in each case study ({\bf \#Reqs}).} 
\label{table:cpsmodels}
\begin{center}
\scalebox{.8}{\begin{tabular}{p{0.15\linewidth}|p{0.82\linewidth} | c | c  }
 \toprule             
 \bf Model Name  & \bf Model Description & {\bf \#Blocks} & {\bf \#Reqs } \\
 \toprule       
	\textsf{Autopilot} & 
A full six degree of freedom simulation of a single-engined high-wing propeller-driven airplane with autopilot. & 1549 & 12 \\           
\midrule
\casestudy  & Discussed in Section~\ref{sec:intro}. & 2192 & 8  \\
	\midrule 
 	\textsf{Neural Network} & A two-input single-output  predictor neural network model with two hidden layers. & 704 & 6  \\ 
 \midrule
\textsf{Tustin} & A numeric model that computes integral over time.& 57 & 5  \\
 	\midrule 	
	\textsf{Regulator} & A typical PID controller.& 308 & 10  \\
 	  	\midrule
  		\textsf{Nonlinear Guidance} &  A non-linear guidance algorithm for an Unmanned Aerial Vehicles (UAV) to follow a moving target.& 373 & 1  \\
	\midrule
	\textsf{System Wide} \newline \textsf{Integrity Monitor} & A numerical algorithm that computes warning to an operator when the airspeed is approaching a boundary where an evasive fly up maneuver cannot be achieved. & 164 & 3   \\
 	\midrule 
	\textsf{Effector Blender} & A control allocation method to calculate the optimal effector configuration for a vehicle. & 95 & 3 \\ 
 	\midrule 
  	\textsf{Two Tanks}  & A two tanks system where a controller regulates the   incoming and outgoing flows of the tanks.  
  	 & 498 & 31  \\
  	\midrule
  	\textsf{Finite State Machine}   & A finite state machine executing in real-time  that turn on the autopilot mode in case of some environment hazard. & 303 & 13  \\
	\midrule 
\textsf{Euler} & A mathematical  model to compute 3-dimensional rotation matrices for an Inertial frame in a Euclidean space.& 834 & 8 \\
  	\bottomrule
\end{tabular}}
\end{center}
\end{table*}

\emph{RQ1 (\ourlogic\ expressiveness).}
To answer this question, we manually formulated the 98 functional requirements in our case studies into the \ourlogic\ language.  All of the 98 functional requirements of our eleven study subjects  were expressible in \ourlogic\ without any need to alter or restrict the requirements descriptions.  Further, all the syntactic constructs of \ourlogic\ described in Section~\ref{sec:FOLfuzzy} were needed to express the requirements in our study. 

\begin{mybox}[colback=lightgray]{}
The answer to RQ1 is that   \ourlogic\ is sufficiently expressive to capture all the 98 CPS requirements of our industrial case studies. 
\end{mybox}

\begin{figure}
\begin{minipage}[c]{\columnwidth}
\centering
\includegraphics[width=1\columnwidth]{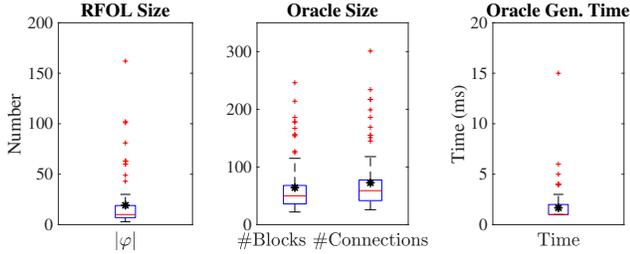}
\caption{Plots reporting (a)~the size of the \ourlogic\ formulas,
(b) the number of blocks and connections of the oracle models
and (c)~the time took SOCRaTEs to generate the oracles.}
\label{fig:SOCRaTeS}
\end{minipage}
\end{figure}

\indent \emph{RQ2 (Usefulness  of the translation).} 
Recall that  engineers need  to write requirements in \ourlogic\  before they can translate them  into Simulink.
To answer this question, we report the size of \ourlogic\ formulas used as input to our approach, the time it takes to generate online Simulink oracles and the size of the generated Simulink oracles. We measure the size of \ourlogic\ requirements  as the sum of the number of quantifiers, and  arithmetic and  logical operators,  and the size of Simulink oracles as their  number of blocks and connections. 
Figure~\ref{fig:SOCRaTeS}(a) shows the size of \ourlogic\ formulas (|$\varphi$|) for our case study requirements,  and Figure~\ref{fig:SOCRaTeS}(b) shows the number of blocks (\#Blocks) and connections  (\#Connections) of the oracle Simulink models that are automatically generated by our approach. In addition, Figure~\ref{fig:SOCRaTeS}(c) shows the time taken by our approach to generate oracle models  from \ourlogic\ formulas. 
As shown in Figure~\ref{fig:SOCRaTeS}, it took on average 1.6ms to automatically generate oracle models with an average number of 64.2 blocks and 72.6 connections for our 98 case study requirements.  Further, the  average size of \ourlogic\ formulas is 19.2, showing that the pre-requisite effort to write the input \ourlogic\   formulas for our approach is not high.
 The difference in size between RFOL formulas and their corresponding Simulink models is mostly due to the former being particularly suitable for expressing declarative properties, such as logical properties with several nested quantifiers. Given this property, and in addition the fact that verification and test engineers are not always very familiar with Simulink — a tool dedicated to control engineers,  we expect significant benefits from translating RFOL into Simulink.

\begin{figure*}[t]
\centering
\hspace{0.5cm}
\begin{minipage}[c]{\textwidth}
\includegraphics[width=.33\columnwidth]{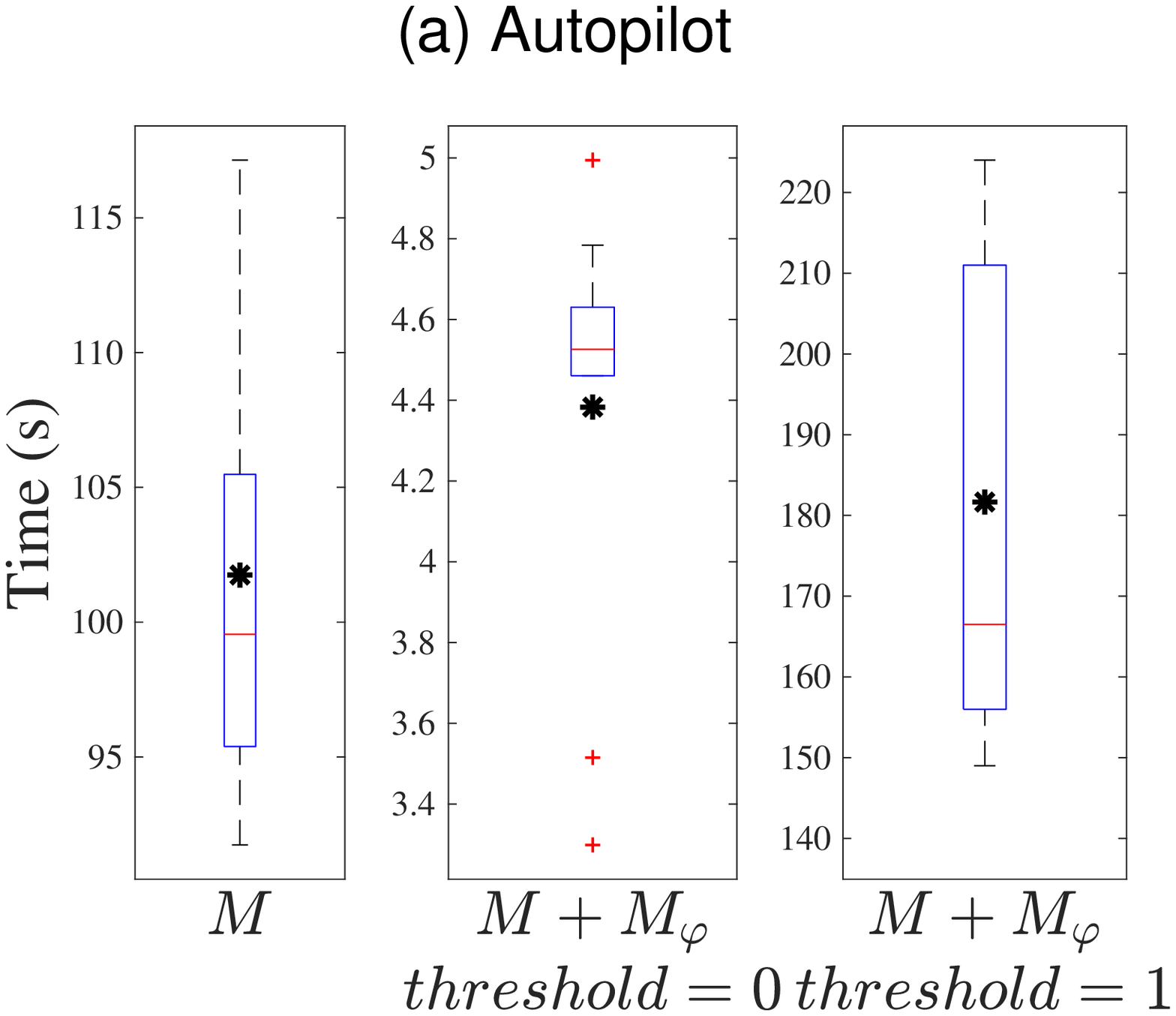}
\includegraphics[width=.33\columnwidth]{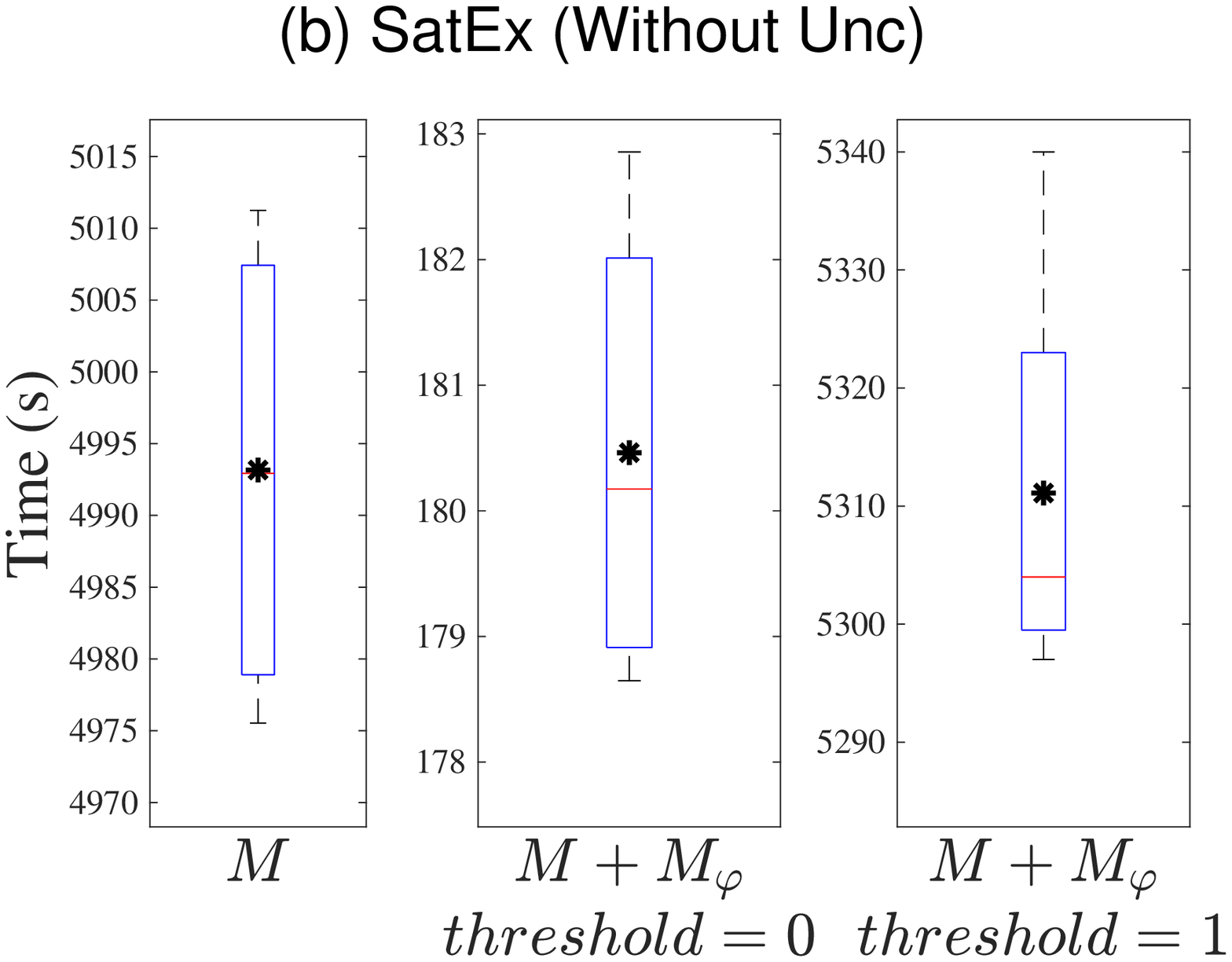}
\includegraphics[width=.33\columnwidth]{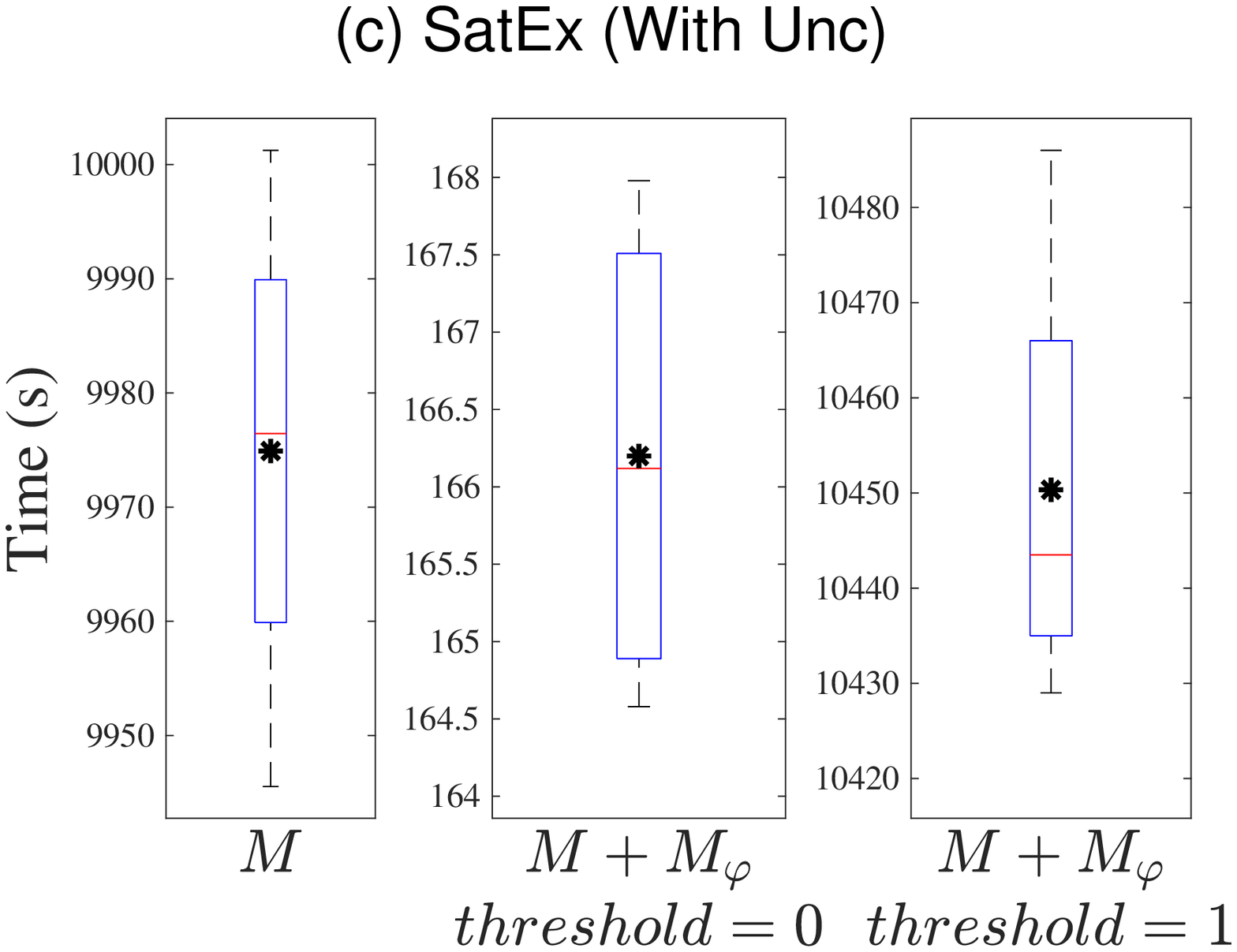}
\caption{Test execution time on models without oracles ($\mathbf{M}$), 
models with oracles ($\mathbf{M + M_\varphi}$) with $\mathbf{\mathbfit{threshold} = 0}$ and models with 
oracles ($\mathbf{M + M_\varphi}$) with $\mathbf{\mathbfit{threshold} = -1}$ for~(a)~\textsf{Autopilot}, (b)~\casestudy\ without uncertainty and (c)~~\casestudy\ with uncertainty.}
\label{fig:rq3}
\end{minipage}
\end{figure*}

\begin{mybox}[colback=lightgray]{}
The answer to RQ2 is that, for our industrial case studies,  the translation into Simulink models is practical as the time required to generate the oracles is acceptable. It takes on average 1.6ms for SOCRaTeS to generate oracle models, and the average size of the input \ourlogic\ formulas is 19.2, showing that the pre-requisite effort of our approach is manageable. 
\end{mybox}

\emph{RQ3 (Impact on the execution time).} Online oracles can save time by stopping test executions  before their completion when they find a failure. However, by combining a model $M$ and a test oracle (i.e., generating $M + M_\varphi$), the model size increases,  and so does its execution time. Hence, in RQ3,  we  compare the time saved by online oracles versus the time overhead of running the oracles together with the models.  For this question, we focus on our two large industrial models, \casestudy\ and \textsf{Autopilot}, since they have long and time-consuming simulations while the other models in Table~\ref{table:cpsmodels} are relatively small with simulation times less than one minute. For such models, both the time savings and the time overheads of our online oracles are  practically insignificant.  

During their internal testing, our partners identified some faults in \casestudy\ and \textsf{Autopilot}  violating some of the model requirements. We received, from our partners, 10 failing test inputs for \textsf{Autopilot} defined over the time  domain $\timedomain = [0,4000]$, and 4 failing test inputs  for \casestudy\ defined over the time domain $\timedomain = [0,86400]$. Recall that  \casestudy\ contains some parameters with uncertain values. We also received the value range for one uncertain parameter of \casestudy, i.e., \textsf{ACM\_type},  from our partner. We then performed the following three experiments. \textbf{EXPI}: We ran all the test inputs on the models alone without including oracle models.  \textbf{EXPII}: We combined  \casestudy\ and \textsf{Autopilot}  with test oracle models related to their respective requirements and ran all the test inputs on the models with oracles. We did not consider any uncertainty in \casestudy\ and set  \textsf{ACM\_type} to a fixed value. \textbf{EXPIII}: We ran  all the tests on \casestudy\ combined with its oracle models and defined   \textsf{ACM\_type} as an uncertain parameter with a value range. We repeated  \textbf{EXPII} and \textbf{EXPIII} for two threshold values: $\mathit{threshold} = 0$ where test executions are stopped when tests fail according to their boolean semantics, and  $\mathit{threshold} = -1$  where test executions are never stopped.

Figures~\ref{fig:rq3}(a) and (b), respectively,  show the results of  \textbf{EXPI} and  \textbf{EXPII} for  \textsf{Autopilot} and \casestudy. Specifically, the figures show the time required to run the test inputs on \textsf{Autopilot} and \casestudy\  (1)~without any oracle model ($M$), (2)~with oracle models ($M+M_\varphi$) for $\mathit{threshold} = 0$, and  (3)~with oracle models ($M+M_\varphi$) for $\mathit{threshold} = -1$. Specifically, in the second case, test oracles stop test executions when test cases fail, and in the third case, test oracles are executed together with the model, but do not stop test executions.  Our results show that on average it takes 101.1s and 4993.2s to run tests on \textsf{Autopilot} and \casestudy, respectively (i.e., Case $M$). These averages, respectively, reduce to 4.3s and 180.4s when oracles stop test executions, and they, respectively, increase to  181.6s and 5311.1s when oracles do not stop  test executions.  That is, for \textsf{Autopilot}, the average time saving of our oracles is 95.6\%  ($\approx$1.5m)  while their average time overhead is 78\% ($\approx$1.2m). In contrast, for \casestudy, our oracles lead to an average time saving of 96\% ($\approx$80m) and an average time overhead of 6\% ($\approx$5m). We note that  \textsf{Autopilot} is less computationally intensive. In this case, the time savings and overheads are almost equivalent because the size and complexity of the generated oracles are comparable to those of the model. \casestudy, on the other hand, is more computationally intensive, and as the results show, for \casestudy\ our oracles  introduce very little time overhead but are able to save a great deal of time when they identify failures. 
Finally, we note that the time saving  depends also on the presence of faults in models and whether and when test cases trigger failures. Nevertheless, according to discussions with our partners, and as evidenced by our case studies, early CPS Simulink models typically contain faults, and hence, our approach can help in saving test execution times for such models.

Figure~\ref{fig:rq3}(c) shows the results of \textbf{EXPIII} for running \casestudy\ with uncertainty. Since in the case of uncertainty, a set of outputs are generated, the total test execution time increases. Specifically, it takes, on average, 9974.9s to run \casestudy\ with uncertainty without oracles, 166.2s to run it  when oracles stop test executions, and  10450.0s to run it when oracles do not stop test executions. As the results show, for \casestudy\ with uncertainty, the time saving is even higher (i.e., 98\%, $\approx$163m) than the case of \casestudy\ without uncertainty, because oracles stop simulations as soon as one output among the set of generated outputs fails.

\begin{mybox}[colback=lightgray]{}
The answer to RQ3 is that,  for large and computationally intensive industrial models, our oracles  introduce very little time overhead (6\%) but are able to save a great deal of time when they identify failures (96\%). When models contain uncertainty the time saving becomes even larger and the time overhead decreases, making our online oracles more beneficial. 
\end{mybox}

\emph{Data Availability.} Our tool is available online~\cite{SOCRaTEs}.
Among the considered models, all the models with the exception of the SatEx model are available on request~\cite{AdditionalMaterial}. 
The SatEx model is not shared as it is part of a non-disclosure agreement.

\section{Related Work}
\label{sec:related}
We classified the related work (Table~\ref{sec:classification}) by analyzing whether the work addresses the oracle generation problem (\textbf{OG})? whether it satisfies assumptions \textbf{A1} and \textbf{A2} (Section~\ref{sec:assumption})? and whether it aims to achieve objectives  \textbf{O1} to  \textbf{O4} (Section~\ref{sec:intro})? Note that assumption \textbf{A3} is considered in all the related work included here. As shown in the table, there is no work that achieves oracle generation and satisfies all our four objectives.  Below, we discuss the closest lines of work to ours among those included in Table~\ref{sec:classification}.

\begin{table}[t]
\caption{Classification of the related work based on the following criteria: Is the work about oracle generation (OG)? Does the work build on the  A1 and A2 assumptions (see Section~\ref{sec:assumption})? Does it achieve the O1 to O4 objectives (see Section~\ref{sec:intro})? Assumption A3 is satisfied by all the related work.
}\label{sec:classification}
\scalebox{.8}{\begin{tabular}{  p{2.5cm} c | c c | c c c c}
\toprule
 & & \multicolumn{2}{c | }{\textbf{Assumpt.}} &\multicolumn{4}{c  }{\textbf{Objectives}} \\
\toprule
\textbf{Ref} & \textbf{OG} & \textbf{A1} &  \textbf{A2} &  \textbf{O1}* &  \textbf{O2} &  \textbf{O3} &  \textbf{O4}  \\
\toprule
\cite{dokhanchi2014line,jakvsic2016quantitative,fainekos2009robustness,donze2010robust} & \cmark & \cmark & \cmark & \xmark$^\prime$ & \cmark & \cmark & \xmark \\
\midrule
\cite{maler2004monitoring,baresi2017test,6732234,bakhirkin2018first},\newline \cite{baresi2017test,6732234,Bartocci:2018:LFS:3178126.3178131} & \cmark & \cmark & \cmark & \xmark  & \cmark &  \xmark & \xmark \\
\midrule
\cite{silvetti2018signal,Maler:2013:MPA:3220915.3221206} & \cmark & \cmark & \cmark & \xmark$^\prime$   & \cmark &  \xmark & \cmark \\
\midrule
 \cite{balsini2017generation} & \cmark & \cmark  & \cmark &  \cmark  & \cmark & \xmark  & \xmark  \\
  \midrule 
  \cite{doi:10.1002/stvr.1464,Dillon:1996:GOY:239098.239116} & \cmark & \cmark & \cmark & \xmark$^\prime$ & \xmark &  \xmark  & \xmark   \\
  \midrule 
 \cite{10.1007/978-3-030-03769-7_20,pill2018automated} & \cmark & \xmark & \xmark &  \xmark$^\prime$ & \xmark & \xmark & \xmark \\
 \midrule
 \cite{hoxha2015vispec,dillon1994specification,Mandrioli:1995:GTC:210223.210226,1521219,10.1007/3-540-61629-2_52,1067980,10.1007/978-3-540-73770-4_6,Morasca:1996:GFT:229000.226300,Stocks:1993:TTS:257572.257664,Richardson:1992:STO:143062.143100,Generating} & \cmark & \xmark & \xmark &  \xmark & \xmark & \xmark & \xmark \\
 \midrule
 \cite{whittle2009relax}  & \xmark & \xmark & \xmark & \xmark & \xmark &  \cmark  & \cmark   \\
\bottomrule 
\end{tabular}}
\flushleft{\hspace*{.cm}\small * 
The notation \xmark$^\prime$ indicates that the monitoring procedure assumes a fixed sample rate, and hence does not accurately handle variable-step outputs.}
\end{table}

Dokhanchi et al.~\cite{dokhanchi2014line} propose an online monitoring procedure for Metric Temporal Logic (MTL)~\cite{koymans1990specifying} properties implemented in the S-TaLiRo tool~\cite{staliro}. The authors use a prediction technique to handle temporal operators that refer to future time instants compared to the  shifting procedures proposed in our work. As a result, their monitoring procedure has a higher running time complexity than our oracles (i.e., polynomial in the size of time history versus linear in the time domain $\timedomain$ size).  Furthermore, they do not translate their monitors into Simulink, and hence,  cannot benefit from the execution time speed-up of efficient Simulink blocks and the Simulink variable step solvers to handle continuous behaviors. 
Thus, as shown by Dokhachi et al.~\cite{dokhanchi2014line}, the time overhead of their approach is considerably high as the time history grows. Jak{\v{s}}i{\'c} et al.~\cite{jakvsic2016quantitative} recently developed an online monitoring procedure for STL by translating  STL into automata monitors with a complexity that is exponential in the size of the formula.  In contrast to our work, such monitors are not able to handle continuous signals sampled at a variable rate directly.
This such signals are approximated as fixed-step signals, hence decreasing the analysis precision of continuous behaviors.  
To the best of our knowledge and according to a recent survey~\cite{bartocci2018specification}, the only work that, like us,  translates a logic into Simulink to enable online monitoring is the work of  Balsini et al.~\cite{balsini2017generation}. The translation, however, is given for a restricted version of STL, which for example does not allow the nesting of more than two temporal operators. As discussed in Section~\ref{sec:FOLfuzzy}, RFOL subsumes STL. Hence, our translation subsumes that of Balsini et al.~\cite{balsini2017generation}.  Breach~\cite{donze2010breach,donze2010robust}   is a monitoring framework for  continuous and hybrid systems that translates STL into  online monitors specified in C++ or MATLAB S-functions. However, due to the overhead of integrating C++ or S-functions in Simulink, running monitors in the Breach framework greatly slows down model simulations, by 4.5 times~\cite{Watanabe:2018:RMS:3195970.3199856}, making the monitors impractical for computationally expensive CPS models  such as our \casestudy\ case study. Finally, Maler et al.~\cite{Maler:2013:MPA:3220915.3221206} propose a monitoring procedure that receives signal segments sequentially, checks each segment and stops simulations if a failure is detected. This work, however, is only partially online since each segment is eventually checked in an offline mode.

There are a number of  work defining a quantitative semantics for logics used in the context of CPS.  Dokhachi et al.~\cite{dokhanchi2014line} introduce the concept of robustness~\cite{fainekos2009robustness} of an STL property specifying how robustly  a requirement is satisfied or violated.  Jak{\v{s}}i{\'c} et al.~\cite{jakvsic2016quantitative} use the concept of edit distance to measure the degree of similarity between two signals.
Silvetti et al.~\cite{silvetti2018signal} propose a variation of STL that enables to reason about the percentage of time a formula is satisfied in a bounded interval. Our semantics for RFOL,  in contrast to the existing work,  captures the satisfaction degree of requirements.

\section{Conclusions}
\label{sec:conclusion}
In this paper, we presented SOCRaTes, an automated approach to generate online test oracles in Simulink able to handle CPS Simulink models with continuous behaviors and involving uncertainties. 
Our oracles generate  a quantitative degree of satisfaction or failure for each test input. 
Our results were obtained by applying SOCRaTes to 11 industry case studies and show that 
(i) our requirements language is able to express all the 98 requirements of our case studies; 
(ii) the effort required by SOCRaTes to generate online oracles in Simulink is acceptable, paving the way for a practical adoption of the approach, and 
(iii) for large models, our approach dramatically reduces the test execution time  compared to when test outputs are checked in an offline manner.  

\bibliographystyle{ACM-Reference-Format}

\begin{thebibliography}{10}

\bibitem{TeXFAQ}
{UK \TeX{} Users Group}.
\newblock {UK} list of {\TeX} frequently asked questions.
\newblock \url{http://www.tex.ac.uk}, 2016.

\bibitem{Downes04:amsart}
Michael Downes and Barbara Beeton.
\newblock {\em The \textsf{amsart}, \textsf{amsproc}, and \textsf{amsbook}
  document~classes}.
\newblock American Mathematical Society, August 2004.
\newblock \url{http://www.ctan.org/pkg/amslatex}.

\bibitem{Fiorio15}
Cristophe Fiorio.
\newblock {\em {a}lgorithm2e.sty---package for algorithms}, October 2015.
\newblock \url{http://www.ctan.org/pkg/algorithm2e}.

\bibitem{Brito09}
Rog\'erio Brito.
\newblock {\em The algorithms bundle}, August 2009.
\newblock \url{http://www.ctan.org/pkg/algorithms}.

\bibitem{Heinz15}
Carsten Heinz, Brooks Moses, and Jobst Hoffmann.
\newblock {\em The Listings Package}, June 2015.
\newblock \url{http://www.ctan.org/pkg/listings}.

\bibitem{Fear05}
Simon Fear.
\newblock {\em Publication quality tables in {\LaTeX}}, April 2005.
\newblock \url{http://www.ctan.org/pkg/booktabs}.

\bibitem{ACMIdentityStandards}
Association for Computing Machinery.
\newblock {\em {ACM} Visual Identity Standards}, 2007.
\newblock \url{http://identitystandards.acm.org}.

\bibitem{Sommerfeldt13:Subcaption}
Axel Sommerfeldt.
\newblock {\em The subcaption package}, April 2013.
\newblock \url{http://www.ctan.org/pkg/subcaption}.

\bibitem{Nomencl}
Boris Veytsman, Bern Schandl, Lee Netherton, and C.~V. Radhakrishnan.
\newblock {\em A package to create a nomenclature}, September 2005.
\newblock \url{http://www.ctan.org/pkg/nomencl}.

\bibitem{Talbot16:Glossaries}
Nicola L.~C. Talbot.
\newblock {\em User Manual for glossaries.sty v4.25}, June 2016.
\newblock \url{http://www.ctan.org/pkg/subcaption}.

\bibitem{Carlisle04:Textcase}
David Carlisle.
\newblock {\em The \textsl{textcase} package}, October 2004.
\newblock \url{http://www.ctan.org/pkg/textcase}.

\end{thebibliography}



\begin{thebibliography}{0}

%%% the Web address.

\ifx \showCODEN    \undefined \def \showCODEN     #1{\unskip}     \fi
\ifx \showDOI      \undefined \def \showDOI       #1{#1}\fi
\ifx \showISBNx    \undefined \def \showISBNx     #1{\unskip}     \fi
\ifx \showISBNxiii \undefined \def \showISBNxiii  #1{\unskip}     \fi
\ifx \showISSN     \undefined \def \showISSN      #1{\unskip}     \fi
\ifx \showLCCN     \undefined \def \showLCCN      #1{\unskip}     \fi
\ifx \shownote     \undefined \def \shownote      #1{#1}          \fi
\ifx \showarticletitle \undefined \def \showarticletitle #1{#1}   \fi
\ifx \showURL      \undefined \def \showURL       {\relax}        \fi
% invisible to TeX
\providecommand\bibfield[2]{#2}
\providecommand\bibinfo[2]{#2}
\providecommand\natexlab[1]{#1}
\providecommand\showeprint[2][]{arXiv:#2}

\end{thebibliography}


\begin{thebibliography}{73}


\ifx \showCODEN    \undefined \def \showCODEN     #1{\unskip}     \fi
\ifx \showDOI      \undefined \def \showDOI       #1{#1}\fi
\ifx \showISBNx    \undefined \def \showISBNx     #1{\unskip}     \fi
\ifx \showISBNxiii \undefined \def \showISBNxiii  #1{\unskip}     \fi
\ifx \showISSN     \undefined \def \showISSN      #1{\unskip}     \fi
\ifx \showLCCN     \undefined \def \showLCCN      #1{\unskip}     \fi
\ifx \shownote     \undefined \def \shownote      #1{#1}          \fi
\ifx \showarticletitle \undefined \def \showarticletitle #1{#1}   \fi
\ifx \showURL      \undefined \def \showURL       {\relax}        \fi
\providecommand\bibfield[2]{#2}
\providecommand\bibinfo[2]{#2}
\providecommand\natexlab[1]{#1}
\providecommand\showeprint[2][]{arXiv:#2}

\bibitem[\protect\citeauthoryear{??}{Lux}{2019}]
        {Luxspace}
 \bibinfo{year}{2019}\natexlab{}.
\newblock \bibinfo{title}{LuxSpace}.
\newblock \bibinfo{howpublished}{https://luxspace.lu/}.
\newblock


\bibitem[\protect\citeauthoryear{??}{mat}{2019}]
        {mathworks}
 \bibinfo{year}{2019}\natexlab{}.
\newblock \bibinfo{title}{{Mathworks}}.
\newblock \bibinfo{howpublished}{\url{https://mathworks.com}}.
\newblock


\bibitem[\protect\citeauthoryear{??}{Add}{2019}]
        {AdditionalMaterial}
 \bibinfo{year}{2019}\natexlab{}.
\newblock \bibinfo{title}{{Online form for accessing companion material}}.
\newblock
  \bibinfo{howpublished}{\newline\url{https://claudiomenghi.github.io/socratesForm.html}}.
\newblock


\bibitem[\protect\citeauthoryear{??}{QRA}{2019}]
        {QRA}
 \bibinfo{year}{2019}\natexlab{}.
\newblock \bibinfo{title}{QRA Corp}.
\newblock \bibinfo{howpublished}{https://qracorp.com/}.
\newblock


\bibitem[\protect\citeauthoryear{??}{qvt}{2019}]
        {qvtrace}
 \bibinfo{year}{2019}\natexlab{}.
\newblock \bibinfo{title}{{QVtrace}}.
\newblock \bibinfo{howpublished}{\url{https://qracorp.com/qvtrace/}}.
\newblock


\bibitem[\protect\citeauthoryear{??}{REA}{2019}]
        {REACTSYS}
 \bibinfo{year}{2019}\natexlab{}.
\newblock \bibinfo{title}{{reactive-systems}}.
\newblock \bibinfo{howpublished}{\url{https://www.reactive-systems.com}}.
\newblock


\bibitem[\protect\citeauthoryear{??}{Rob}{2019}]
        {RobustControlToolbox}
 \bibinfo{year}{2019}\natexlab{}.
\newblock \bibinfo{title}{Robust Control Toolbox}.
\newblock
  \bibinfo{howpublished}{\url{https://nl.mathworks.com/products/robust.html}}.
\newblock


\bibitem[\protect\citeauthoryear{??}{Sig}{2019}]
        {SignalToNoise}
 \bibinfo{year}{2019}\natexlab{}.
\newblock \bibinfo{title}{{Signal To Noise Ratio}}.
\newblock
  \bibinfo{howpublished}{\url{https://en.wikipedia.org/wiki/Signal-to-noise_ratio}}.
\newblock


\bibitem[\protect\citeauthoryear{??}{SDV}{2019}]
        {SDV}
 \bibinfo{year}{2019}\natexlab{}.
\newblock \bibinfo{title}{{Simulink Design Verifier}}.
\newblock
  \bibinfo{howpublished}{\url{https://nl.mathworks.com/products/sldesignverifier.html}}.
\newblock


\bibitem[\protect\citeauthoryear{??}{Sim}{2019}]
        {SimulinkSolvers}
 \bibinfo{year}{2019}\natexlab{}.
\newblock \bibinfo{title}{Simulink Solvers}.
\newblock
  \bibinfo{howpublished}{\url{https://nl.mathworks.com/help/simulink/ug/types-of-solvers.html}}.
\newblock


\bibitem[\protect\citeauthoryear{??}{sir}{2019}]
        {sirius}
 \bibinfo{year}{2019}\natexlab{}.
\newblock \bibinfo{title}{Sirius}.
\newblock \bibinfo{howpublished}{\url{https://www.eclipse.org/sirius/}}.
\newblock


\bibitem[\protect\citeauthoryear{??}{SOC}{2019}]
        {SOCRaTEs}
 \bibinfo{year}{2019}\natexlab{}.
\newblock \bibinfo{title}{SOCRaTEs}.
\newblock \bibinfo{howpublished}{https://github.com/claudiomenghi/SOCRaTEs}.
\newblock


\bibitem[\protect\citeauthoryear{??}{ure}{2019}]
        {ureal}
 \bibinfo{year}{2019}\natexlab{}.
\newblock \bibinfo{title}{Uncertain Real Parameters}.
\newblock
  \bibinfo{howpublished}{\url{https://nl.mathworks.com/help/robust/ug/uncertain-real-parameters.html}}.
\newblock


\bibitem[\protect\citeauthoryear{??}{Xte}{2019}]
        {Xtext}
 \bibinfo{year}{2019}\natexlab{}.
\newblock \bibinfo{title}{Xtext}.
\newblock \bibinfo{howpublished}{\url{https://www.eclipse.org/Xtext/}}.
\newblock


\bibitem[\protect\citeauthoryear{Abbas, Fainekos, Sankaranarayanan, Ivancic,
  and Gupta}{Abbas et~al\mbox{.}}{2013}]
        {Abbas:13}
\bibfield{author}{\bibinfo{person}{Houssam Abbas}, \bibinfo{person}{Georgios~E.
  Fainekos}, \bibinfo{person}{Sriram Sankaranarayanan}, \bibinfo{person}{Franjo
  Ivancic}, {and} \bibinfo{person}{Aarti Gupta}.}
  \bibinfo{year}{2013}\natexlab{}.
\newblock \showarticletitle{Probabilistic Temporal Logic Falsification of
  Cyber-Physical Systems}.
\newblock \bibinfo{journal}{\emph{{ACM} Transactions on Embedded Computing
  Systems (TECS)}} \bibinfo{volume}{12}, \bibinfo{number}{2s}
  (\bibinfo{year}{2013}), \bibinfo{pages}{95:1--95:30}.
\newblock


\bibitem[\protect\citeauthoryear{{Alur}}{{Alur}}{2011}]
        {6064535}
\bibfield{author}{\bibinfo{person}{R. {Alur}}.}
  \bibinfo{year}{2011}\natexlab{}.
\newblock \showarticletitle{Formal verification of hybrid systems}. In
  \bibinfo{booktitle}{\emph{International Conference on Embedded Software
  (EMSOFT)}}. \bibinfo{pages}{273--278}.
\newblock


\bibitem[\protect\citeauthoryear{Alur}{Alur}{2015}]
        {alur:15}
\bibfield{author}{\bibinfo{person}{Rajeev Alur}.}
  \bibinfo{year}{2015}\natexlab{}.
\newblock \bibinfo{booktitle}{\emph{Principles of Cyber-Physical Systems}}.
\newblock \bibinfo{publisher}{MIT Press}.
\newblock


\bibitem[\protect\citeauthoryear{Alur, Courcoubetis, Halbwachs, Henzinger, Ho,
  Nicollin, Olivero, Sifakis, and Yovine}{Alur et~al\mbox{.}}{1995}]
        {alur1995algorithmic}
\bibfield{author}{\bibinfo{person}{Rajeev Alur}, \bibinfo{person}{Costas
  Courcoubetis}, \bibinfo{person}{Nicolas Halbwachs}, \bibinfo{person}{Thomas~A
  Henzinger}, \bibinfo{person}{P-H Ho}, \bibinfo{person}{Xavier Nicollin},
  \bibinfo{person}{Alfredo Olivero}, \bibinfo{person}{Joseph Sifakis}, {and}
  \bibinfo{person}{Sergio Yovine}.} \bibinfo{year}{1995}\natexlab{}.
\newblock \showarticletitle{The algorithmic analysis of hybrid systems}.
\newblock \bibinfo{journal}{\emph{Theoretical computer science}}
  \bibinfo{volume}{138}, \bibinfo{number}{1} (\bibinfo{year}{1995}),
  \bibinfo{pages}{3--34}.
\newblock


\bibitem[\protect\citeauthoryear{Andrés, Merayo, and Núñez}{Andrés
  et~al\mbox{.}}{2012}]
        {doi:10.1002/stvr.1464}
\bibfield{author}{\bibinfo{person}{César Andrés},
  \bibinfo{person}{Mercedes~G. Merayo}, {and} \bibinfo{person}{Manuel
  Núñez}.} \bibinfo{year}{2012}\natexlab{}.
\newblock \showarticletitle{Formal passive testing of timed systems: theory and
  tools}.
\newblock \bibinfo{journal}{\emph{Software Testing, Verification and
  Reliability}} \bibinfo{volume}{22}, \bibinfo{number}{6}
  (\bibinfo{year}{2012}), \bibinfo{pages}{365--405}.
\newblock
\urldef\tempurl
\url{https://doi.org/10.1002/stvr.1464}
\showDOI{\tempurl}


\bibitem[\protect\citeauthoryear{Annpureddy, Liu, Fainekos, and
  Sankaranarayanan}{Annpureddy et~al\mbox{.}}{2011}]
        {staliro}
\bibfield{author}{\bibinfo{person}{Yashwanth Annpureddy}, \bibinfo{person}{Che
  Liu}, \bibinfo{person}{Georgios Fainekos}, {and} \bibinfo{person}{Sriram
  Sankaranarayanan}.} \bibinfo{year}{2011}\natexlab{}.
\newblock \showarticletitle{S-TaLiRo: A Tool for Temporal Logic Falsification
  for Hybrid Systems}. In \bibinfo{booktitle}{\emph{Tools and Algorithms for
  the Construction and Analysis of Systems}},
  \bibfield{editor}{\bibinfo{person}{Parosh~Aziz Abdulla} {and}
  \bibinfo{person}{K.~Rustan~M. Leino}} (Eds.). \bibinfo{publisher}{Springer},
  \bibinfo{pages}{254--257}.
\newblock


\bibitem[\protect\citeauthoryear{Atkinson}{Atkinson}{2008}]
        {atkinson2008introduction}
\bibfield{author}{\bibinfo{person}{Kendall~E Atkinson}.}
  \bibinfo{year}{2008}\natexlab{}.
\newblock \bibinfo{booktitle}{\emph{An introduction to numerical analysis}}.
\newblock \bibinfo{publisher}{John Wiley \& Sons}.
\newblock


\bibitem[\protect\citeauthoryear{Bakhirkin, Ferr{\`e}re, Henzinger, and
  Ni{\v{c}}kovi{\'c}}{Bakhirkin et~al\mbox{.}}{2018a}]
        {bakhirkin2018first}
\bibfield{author}{\bibinfo{person}{Alexey Bakhirkin}, \bibinfo{person}{Thomas
  Ferr{\`e}re}, \bibinfo{person}{Thomas~A Henzinger}, {and}
  \bibinfo{person}{Dejan Ni{\v{c}}kovi{\'c}}.}
  \bibinfo{year}{2018}\natexlab{a}.
\newblock \showarticletitle{The first-order logic of signals: keynote}. In
  \bibinfo{booktitle}{\emph{International Conference on Embedded Software}}.
  IEEE Press, \bibinfo{pages}{1}.
\newblock


\bibitem[\protect\citeauthoryear{Bakhirkin, Ferr\`{e}re, Henzinger, and
  Ni\v{c}kovi\'{c}}{Bakhirkin et~al\mbox{.}}{2018b}]
        {Bakhirkin:2018:FLS:3283535.3283536}
\bibfield{author}{\bibinfo{person}{Alexey Bakhirkin}, \bibinfo{person}{Thomas
  Ferr\`{e}re}, \bibinfo{person}{Thomas~A. Henzinger}, {and}
  \bibinfo{person}{Dejan Ni\v{c}kovi\'{c}}.} \bibinfo{year}{2018}\natexlab{b}.
\newblock \showarticletitle{The First-order Logic of Signals: Keynote}. In
  \bibinfo{booktitle}{\emph{International Conference on Embedded Software}}
  \emph{(\bibinfo{series}{EMSOFT})}. \bibinfo{publisher}{IEEE}, Article
  \bibinfo{articleno}{1}, \bibinfo{numpages}{10}~pages.
\newblock
\showISBNx{978-1-5386-5564-1}
\urldef\tempurl
\url{http://dl.acm.org/citation.cfm?id=3283535.3283536}
\showURL{
\tempurl}


\bibitem[\protect\citeauthoryear{Balsini, Di~Natale, Celia, and
  Tsachouridis}{Balsini et~al\mbox{.}}{2017}]
        {balsini2017generation}
\bibfield{author}{\bibinfo{person}{Alessio Balsini}, \bibinfo{person}{Marco
  Di~Natale}, \bibinfo{person}{Marco Celia}, {and} \bibinfo{person}{Vassilios
  Tsachouridis}.} \bibinfo{year}{2017}\natexlab{}.
\newblock \showarticletitle{Generation of Simulink monitors for control
  applications from formal requirements}. In
  \bibinfo{booktitle}{\emph{Industrial Embedded Systems (SIES)}}. IEEE,
  \bibinfo{pages}{1--9}.
\newblock


\bibitem[\protect\citeauthoryear{Baresi, Delamaro, and Nardi}{Baresi
  et~al\mbox{.}}{2017}]
        {baresi2017test}
\bibfield{author}{\bibinfo{person}{Luciano Baresi}, \bibinfo{person}{Marcio
  Delamaro}, {and} \bibinfo{person}{Paulo Nardi}.}
  \bibinfo{year}{2017}\natexlab{}.
\newblock \showarticletitle{Test oracles for simulink-like models}.
\newblock \bibinfo{journal}{\emph{Automated Software Engineering}}
  \bibinfo{volume}{24}, \bibinfo{number}{2} (\bibinfo{year}{2017}),
  \bibinfo{pages}{369--391}.
\newblock


\bibitem[\protect\citeauthoryear{Barr, Harman, McMinn, Shahbaz, and Yoo}{Barr
  et~al\mbox{.}}{2015}]
        {barr2015oracle}
\bibfield{author}{\bibinfo{person}{Earl~T Barr}, \bibinfo{person}{Mark Harman},
  \bibinfo{person}{Phil McMinn}, \bibinfo{person}{Muzammil Shahbaz}, {and}
  \bibinfo{person}{Shin Yoo}.} \bibinfo{year}{2015}\natexlab{}.
\newblock \showarticletitle{The oracle problem in software testing: A survey}.
\newblock \bibinfo{journal}{\emph{Transactions on Software Engineering}}
  \bibinfo{volume}{41}, \bibinfo{number}{5} (\bibinfo{year}{2015}),
  \bibinfo{pages}{507--525}.
\newblock


\bibitem[\protect\citeauthoryear{Bartocci, Deshmukh, Donz{\'e}, Fainekos,
  Maler, Ni{\v{c}}kovi{\'c}, and Sankaranarayanan}{Bartocci
  et~al\mbox{.}}{2018a}]
        {bartocci2018specification}
\bibfield{author}{\bibinfo{person}{Ezio Bartocci}, \bibinfo{person}{Jyotirmoy
  Deshmukh}, \bibinfo{person}{Alexandre Donz{\'e}}, \bibinfo{person}{Georgios
  Fainekos}, \bibinfo{person}{Oded Maler}, \bibinfo{person}{Dejan
  Ni{\v{c}}kovi{\'c}}, {and} \bibinfo{person}{Sriram Sankaranarayanan}.}
  \bibinfo{year}{2018}\natexlab{a}.
\newblock \showarticletitle{Specification-based monitoring of cyber-physical
  systems: a survey on theory, tools and applications}.
\newblock In \bibinfo{booktitle}{\emph{Lectures on Runtime Verification}}.
  \bibinfo{publisher}{Springer}, \bibinfo{pages}{135--175}.
\newblock


\bibitem[\protect\citeauthoryear{Bartocci, Ferr\`{e}re, Manjunath, and
  Ni\v{c}kovi\'{c}}{Bartocci et~al\mbox{.}}{2018b}]
        {Bartocci:2018:LFS:3178126.3178131}
\bibfield{author}{\bibinfo{person}{Ezio Bartocci}, \bibinfo{person}{Thomas
  Ferr\`{e}re}, \bibinfo{person}{Niveditha Manjunath}, {and}
  \bibinfo{person}{Dejan Ni\v{c}kovi\'{c}}.} \bibinfo{year}{2018}\natexlab{b}.
\newblock \showarticletitle{Localizing Faults in Simulink/Stateflow Models with
  STL}. In \bibinfo{booktitle}{\emph{International Conference on Hybrid
  Systems: Computation and Control}} \emph{(\bibinfo{series}{HSCC})}.
  \bibinfo{publisher}{ACM}, \bibinfo{pages}{197--206}.
\newblock
\showISBNx{978-1-4503-5642-8}


\bibitem[\protect\citeauthoryear{Coppit and Haddox-Schatz}{Coppit and
  Haddox-Schatz}{2005}]
        {1521219}
\bibfield{author}{\bibinfo{person}{D. Coppit} {and} \bibinfo{person}{J.~M.
  Haddox-Schatz}.} \bibinfo{year}{2005}\natexlab{}.
\newblock \showarticletitle{On the Use of Specification-Based Assertions as
  Test Oracles}. In \bibinfo{booktitle}{\emph{NASA Software Engineering
  Workshop}}. \bibinfo{publisher}{IEEE}, \bibinfo{pages}{305--314}.
\newblock
\showISSN{1550-6215}


\bibitem[\protect\citeauthoryear{de~Weck~Olivier, John,
  et~al\mbox{.}}{de~Weck~Olivier et~al\mbox{.}}{2007}]
        {de2007classification}
\bibfield{author}{\bibinfo{person}{Eckert Claudia~M de Weck~Olivier},
  \bibinfo{person}{Clarkson~P John}, {et~al\mbox{.}}}
  \bibinfo{year}{2007}\natexlab{}.
\newblock \showarticletitle{A classification of uncertainty for early product
  and system design}.
\newblock \bibinfo{journal}{\emph{Guidelines for a Decision Support Method
  Adapted to NPD Processes}} (\bibinfo{year}{2007}), \bibinfo{pages}{159--160}.
\newblock


\bibitem[\protect\citeauthoryear{Dillon and Ramakrishna}{Dillon and
  Ramakrishna}{1996}]
        {Dillon:1996:GOY:239098.239116}
\bibfield{author}{\bibinfo{person}{L.~K. Dillon} {and} \bibinfo{person}{Y.~S.
  Ramakrishna}.} \bibinfo{year}{1996}\natexlab{}.
\newblock \showarticletitle{Generating Oracles from Your Favorite Temporal
  Logic Specifications}. In \bibinfo{booktitle}{\emph{Symposium on Foundations
  of Software Engineering}} \emph{(\bibinfo{series}{SIGSOFT})}.
  \bibinfo{publisher}{ACM}.
\newblock


\bibitem[\protect\citeauthoryear{Dillon and Yu}{Dillon and Yu}{1994}]
        {dillon1994specification}
\bibfield{author}{\bibinfo{person}{Laura~K Dillon} {and} \bibinfo{person}{Qing
  Yu}.} \bibinfo{year}{1994}\natexlab{}.
\newblock \showarticletitle{Specification and Testing of Temporal Properties of
  Concurrent System Designs}. University of California at Santa Barbara.
\newblock


\bibitem[\protect\citeauthoryear{Dokhanchi, Hoxha, and Fainekos}{Dokhanchi
  et~al\mbox{.}}{2014}]
        {dokhanchi2014line}
\bibfield{author}{\bibinfo{person}{Adel Dokhanchi}, \bibinfo{person}{Bardh
  Hoxha}, {and} \bibinfo{person}{Georgios Fainekos}.}
  \bibinfo{year}{2014}\natexlab{}.
\newblock \showarticletitle{On-line monitoring for temporal logic robustness}.
  In \bibinfo{booktitle}{\emph{International Conference on Runtime
  Verification}}. Springer, \bibinfo{pages}{231--246}.
\newblock


\bibitem[\protect\citeauthoryear{Donz{\'e}}{Donz{\'e}}{2010}]
        {donze2010breach}
\bibfield{author}{\bibinfo{person}{Alexandre Donz{\'e}}.}
  \bibinfo{year}{2010}\natexlab{}.
\newblock \showarticletitle{Breach, a toolbox for verification and parameter
  synthesis of hybrid systems}. In \bibinfo{booktitle}{\emph{Computer Aided
  Verification}}. Springer, \bibinfo{pages}{167--170}.
\newblock


\bibitem[\protect\citeauthoryear{Donz{\'e} and Maler}{Donz{\'e} and
  Maler}{2010}]
        {donze2010robust}
\bibfield{author}{\bibinfo{person}{Alexandre Donz{\'e}} {and}
  \bibinfo{person}{Oded Maler}.} \bibinfo{year}{2010}\natexlab{}.
\newblock \showarticletitle{Robust satisfaction of temporal logic over
  real-valued signals}. In \bibinfo{booktitle}{\emph{International Conference
  on Formal Modeling and Analysis of Timed Systems}}. Springer,
  \bibinfo{pages}{92--106}.
\newblock


\bibitem[\protect\citeauthoryear{Elbaum and Rosenblum}{Elbaum and
  Rosenblum}{[n. d.]}]
        {elbaum2014known}
\bibfield{author}{\bibinfo{person}{Sebastian Elbaum} {and}
  \bibinfo{person}{David~S Rosenblum}.} \bibinfo{year}{[n. d.]}\natexlab{}.
\newblock \showarticletitle{Known unknowns: testing in the presence of
  uncertainty}.
\newblock


\bibitem[\protect\citeauthoryear{Fainekos and Pappas}{Fainekos and
  Pappas}{2009}]
        {fainekos2009robustness}
\bibfield{author}{\bibinfo{person}{Georgios~E Fainekos} {and}
  \bibinfo{person}{George~J Pappas}.} \bibinfo{year}{2009}\natexlab{}.
\newblock \showarticletitle{Robustness of temporal logic specifications for
  continuous-time signals}.
\newblock \bibinfo{journal}{\emph{Theoretical Computer Science}}
  \bibinfo{volume}{410}, \bibinfo{number}{42} (\bibinfo{year}{2009}),
  \bibinfo{pages}{4262--4291}.
\newblock


\bibitem[\protect\citeauthoryear{Golnaraghi and Kuo}{Golnaraghi and
  Kuo}{2010}]
        {golnaraghi2010automatic}
\bibfield{author}{\bibinfo{person}{Farid Golnaraghi} {and} \bibinfo{person}{BC
  Kuo}.} \bibinfo{year}{2010}\natexlab{}.
\newblock \showarticletitle{Automatic control systems}.
\newblock \bibinfo{journal}{\emph{Complex Variables}}  \bibinfo{volume}{2}
  (\bibinfo{year}{2010}), \bibinfo{pages}{1--1}.
\newblock


\bibitem[\protect\citeauthoryear{Harman, McMinn, Shahbaz, and Yoo}{Harman
  et~al\mbox{.}}{2013}]
        {harman2013comprehensive}
\bibfield{author}{\bibinfo{person}{Mark Harman}, \bibinfo{person}{Phil McMinn},
  \bibinfo{person}{Muzammil Shahbaz}, {and} \bibinfo{person}{Shin Yoo}.}
  \bibinfo{year}{2013}\natexlab{}.
\newblock \showarticletitle{A comprehensive survey of trends in oracles for
  software testing}.
\newblock \bibinfo{journal}{\emph{University of Sheffield, Department of
  Computer Science, Tech. Rep. CS-13-01}} (\bibinfo{year}{2013}).
\newblock


\bibitem[\protect\citeauthoryear{Henzinger, Kopke, Puri, and Varaiya}{Henzinger
  et~al\mbox{.}}{1998}]
        {henzinger1998s}
\bibfield{author}{\bibinfo{person}{Thomas~A Henzinger},
  \bibinfo{person}{Peter~W Kopke}, \bibinfo{person}{Anuj Puri}, {and}
  \bibinfo{person}{Pravin Varaiya}.} \bibinfo{year}{1998}\natexlab{}.
\newblock \showarticletitle{What's decidable about hybrid automata?}
\newblock \bibinfo{journal}{\emph{Journal of computer and system sciences}}
  \bibinfo{volume}{57}, \bibinfo{number}{1} (\bibinfo{year}{1998}),
  \bibinfo{pages}{94--124}.
\newblock


\bibitem[\protect\citeauthoryear{Hoxha, Mavridis, and Fainekos}{Hoxha
  et~al\mbox{.}}{2015}]
        {hoxha2015vispec}
\bibfield{author}{\bibinfo{person}{Bardh Hoxha}, \bibinfo{person}{Nikolaos
  Mavridis}, {and} \bibinfo{person}{Georgios Fainekos}.}
  \bibinfo{year}{2015}\natexlab{}.
\newblock \showarticletitle{VISPEC: A graphical tool for elicitation of MTL
  requirements}. In \bibinfo{booktitle}{\emph{Intelligent Robots and Systems
  (IROS)}}. IEEE, \bibinfo{pages}{3486--3492}.
\newblock


\bibitem[\protect\citeauthoryear{IEC 61508 (2010)}{IEC 61508 (2010)}{[n. d.]}]
        {IEC61508}
IEC 61508 (2010) \bibinfo{year}{[n. d.]}\natexlab{}.
\newblock \bibinfo{booktitle}{\emph{Functional safety of
  electrical/electronic/programmable electronic safety-related systems}
  (\bibinfo{edition}{2.0} ed.)}.
\newblock \bibinfo{type}{International Standard}. \bibinfo{institution}{IEC}.
\newblock


\bibitem[\protect\citeauthoryear{Jak{\v{s}}i{\'c}, Bartocci, Grosu, and
  Ni{\v{c}}kovi{\'c}}{Jak{\v{s}}i{\'c} et~al\mbox{.}}{2016}]
        {jakvsic2016quantitative}
\bibfield{author}{\bibinfo{person}{Stefan Jak{\v{s}}i{\'c}},
  \bibinfo{person}{Ezio Bartocci}, \bibinfo{person}{Radu Grosu}, {and}
  \bibinfo{person}{Dejan Ni{\v{c}}kovi{\'c}}.} \bibinfo{year}{2016}\natexlab{}.
\newblock \showarticletitle{Quantitative monitoring of STL with edit distance}.
  In \bibinfo{booktitle}{\emph{International Conference on Runtime
  Verification}}. Springer, \bibinfo{pages}{201--218}.
\newblock


\bibitem[\protect\citeauthoryear{Kent, Holzman, and Belzer}{Kent
  et~al\mbox{.}}{1975}]
        {FSA}
\bibfield{author}{\bibinfo{person}{Allen Kent}, \bibinfo{person}{Albert~George
  Holzman}, {and} \bibinfo{person}{Jack Belzer}.}
  \bibinfo{year}{1975}\natexlab{}.
\newblock \showarticletitle{Encyclopedia of computer science and technology}.
\newblock  (\bibinfo{year}{1975}).
\newblock


\bibitem[\protect\citeauthoryear{Koymans}{Koymans}{1990}]
        {koymans1990specifying}
\bibfield{author}{\bibinfo{person}{Ron Koymans}.}
  \bibinfo{year}{1990}\natexlab{}.
\newblock \showarticletitle{Specifying real-time properties with metric
  temporal logic}.
\newblock \bibinfo{journal}{\emph{Real-time systems}} \bibinfo{volume}{2},
  \bibinfo{number}{4} (\bibinfo{year}{1990}), \bibinfo{pages}{255--299}.
\newblock


\bibitem[\protect\citeauthoryear{Larsen and Thomsen}{Larsen and
  Thomsen}{1988}]
        {larsen1988modal}
\bibfield{author}{\bibinfo{person}{Kim~G Larsen} {and} \bibinfo{person}{Bent
  Thomsen}.} \bibinfo{year}{1988}\natexlab{}.
\newblock \showarticletitle{A modal process logic}. In
  \bibinfo{booktitle}{\emph{Logic in Computer Science}}. IEEE,
  \bibinfo{pages}{203--210}.
\newblock


\bibitem[\protect\citeauthoryear{Le~Gall and Arnould}{Le~Gall and
  Arnould}{1996}]
        {10.1007/3-540-61629-2_52}
\bibfield{author}{\bibinfo{person}{Pascale Le~Gall} {and}
  \bibinfo{person}{Agn{\`e}s Arnould}.} \bibinfo{year}{1996}\natexlab{}.
\newblock \showarticletitle{Formal specifications and test: Correctness and
  oracle}. In \bibinfo{booktitle}{\emph{Recent Trends in Data Type
  Specification}}, \bibfield{editor}{\bibinfo{person}{Magne Haveraaen},
  \bibinfo{person}{Olaf Owe}, {and} \bibinfo{person}{Ole-Johan Dahl}} (Eds.).
  \bibinfo{publisher}{Springer}.
\newblock


\bibitem[\protect\citeauthoryear{Lin and Ho}{Lin and Ho}{2001}]
        {Generating}
\bibfield{author}{\bibinfo{person}{Jin-Cherng Lin} {and} \bibinfo{person}{Ian
  Ho}.} \bibinfo{year}{2001}\natexlab{}.
\newblock \showarticletitle{Generating timed test cases with oracles for
  real-time software}.
\newblock \bibinfo{journal}{\emph{Advances in Engineering Software}}
  \bibinfo{volume}{32}, \bibinfo{number}{9} (\bibinfo{year}{2001}),
  \bibinfo{pages}{705 -- 715}.
\newblock
\showISSN{0965-9978}
\urldef\tempurl
\url{https://doi.org/10.1016/S0965-9978(01)00021-7}
\showDOI{\tempurl}


\bibitem[\protect\citeauthoryear{Maler and Nickovic}{Maler and
  Nickovic}{2004}]
        {maler2004monitoring}
\bibfield{author}{\bibinfo{person}{Oded Maler} {and} \bibinfo{person}{Dejan
  Nickovic}.} \bibinfo{year}{2004}\natexlab{}.
\newblock \showarticletitle{Monitoring temporal properties of continuous
  signals}.
\newblock In \bibinfo{booktitle}{\emph{Formal Techniques, Modelling and
  Analysis of Timed and Fault-Tolerant Systems}}.
  \bibinfo{publisher}{Springer}, \bibinfo{pages}{152--166}.
\newblock


\bibitem[\protect\citeauthoryear{Maler and Ni\u{a}\'{z}Kovi\'{c}}{Maler and
  Ni\u{a}\'{z}Kovi\'{c}}{2013}]
        {Maler:2013:MPA:3220915.3221206}
\bibfield{author}{\bibinfo{person}{Oded Maler} {and} \bibinfo{person}{Dejan
  Ni\u{a}\'{z}Kovi\'{c}}.} \bibinfo{year}{2013}\natexlab{}.
\newblock \showarticletitle{Monitoring Properties of Analog and Mixed-signal
  Circuits}.
\newblock \bibinfo{journal}{\emph{Int. J. Softw. Tools Technol. Transf.}}
  \bibinfo{volume}{15}, \bibinfo{number}{3} (\bibinfo{date}{June}
  \bibinfo{year}{2013}), \bibinfo{pages}{247--268}.
\newblock
\showISSN{1433-2779}
\urldef\tempurl
\url{https://doi.org/10.1007/s10009-012-0247-9}
\showDOI{\tempurl}


\bibitem[\protect\citeauthoryear{Mandrioli, Morasca, and Morzenti}{Mandrioli
  et~al\mbox{.}}{1995}]
        {Mandrioli:1995:GTC:210223.210226}
\bibfield{author}{\bibinfo{person}{Dino Mandrioli}, \bibinfo{person}{Sandro
  Morasca}, {and} \bibinfo{person}{Angelo Morzenti}.}
  \bibinfo{year}{1995}\natexlab{}.
\newblock \showarticletitle{Generating Test Cases for Real-time Systems from
  Logic Specifications}.
\newblock \bibinfo{journal}{\emph{ACM Trans. Comput. Syst.}}
  \bibinfo{volume}{13}, \bibinfo{number}{4} (\bibinfo{date}{Nov.}
  \bibinfo{year}{1995}), \bibinfo{pages}{365--398}.
\newblock
\showISSN{0734-2071}
\urldef\tempurl
\url{https://doi.org/10.1145/210223.210226}
\showDOI{\tempurl}


\bibitem[\protect\citeauthoryear{Matinnejad, Nejati, Briand, and
  Bruckmann}{Matinnejad et~al\mbox{.}}{2018}]
        {matinnejad2018test}
\bibfield{author}{\bibinfo{person}{Reza Matinnejad}, \bibinfo{person}{Shiva
  Nejati}, \bibinfo{person}{Lionel Briand}, {and} \bibinfo{person}{Thomas
  Bruckmann}.} \bibinfo{year}{2018}\natexlab{}.
\newblock \showarticletitle{Test Generation and Test Prioritization for
  Simulink Models with Dynamic Behavior}.
\newblock \bibinfo{journal}{\emph{IEEE Transactions on Software Engineering}}
  (\bibinfo{year}{2018}).
\newblock


\bibitem[\protect\citeauthoryear{Matinnejad, Nejati, Briand, and
  Bruckmann}{Matinnejad et~al\mbox{.}}{2016}]
        {Matinnejad:2016:ATS:2884781.2884797}
\bibfield{author}{\bibinfo{person}{Reza Matinnejad}, \bibinfo{person}{Shiva
  Nejati}, \bibinfo{person}{Lionel~C. Briand}, {and} \bibinfo{person}{Thomas
  Bruckmann}.} \bibinfo{year}{2016}\natexlab{}.
\newblock \showarticletitle{Automated Test Suite Generation for Time-continuous
  Simulink Models}. In \bibinfo{booktitle}{\emph{International Conference on
  Software Engineering}} \emph{(\bibinfo{series}{ICSE})}.
  \bibinfo{publisher}{ACM}.
\newblock


\bibitem[\protect\citeauthoryear{Moiz}{Moiz}{2017}]
        {moiz2017uncertainty}
\bibfield{author}{\bibinfo{person}{Salman~Abdul Moiz}.}
  \bibinfo{year}{2017}\natexlab{}.
\newblock \showarticletitle{Uncertainty in Software Testing}.
\newblock In \bibinfo{booktitle}{\emph{Trends in Software Testing}}.
  \bibinfo{publisher}{Springer}, \bibinfo{pages}{67--87}.
\newblock


\bibitem[\protect\citeauthoryear{Morasca, Morzenti, and SanPietro}{Morasca
  et~al\mbox{.}}{1996}]
        {Morasca:1996:GFT:229000.226300}
\bibfield{author}{\bibinfo{person}{Sandro Morasca}, \bibinfo{person}{Angelo
  Morzenti}, {and} \bibinfo{person}{Pieluigi SanPietro}.}
  \bibinfo{year}{1996}\natexlab{}.
\newblock \showarticletitle{Generating Functional Test Cases In-the-large for
  Time-critical Systems from Logic-based Specifications}. In
  \bibinfo{booktitle}{\emph{1996 ACM SIGSOFT International Symposium on
  Software Testing and Analysis}} \emph{(\bibinfo{series}{ISSTA '96})}.
  \bibinfo{publisher}{ACM}, \bibinfo{address}{New York, NY, USA},
  \bibinfo{pages}{39--52}.
\newblock
\showISBNx{0-89791-787-1}
\urldef\tempurl
\url{https://doi.org/10.1145/229000.226300}
\showDOI{\tempurl}


\bibitem[\protect\citeauthoryear{Nardi and Damasceno}{Nardi and
  Damasceno}{2015}]
        {nardisurvey}
\bibfield{author}{\bibinfo{person}{Paulo~A Nardi} {and}
  \bibinfo{person}{Eduardo~F Damasceno}.} \bibinfo{year}{2015}\natexlab{}.
\newblock \showarticletitle{A Survey on Test Oracles}.
\newblock  (\bibinfo{year}{2015}).
\newblock


\bibitem[\protect\citeauthoryear{Nardi, Delamaro, and Baresi}{Nardi
  et~al\mbox{.}}{2013}]
        {6732234}
\bibfield{author}{\bibinfo{person}{P.~A. Nardi}, \bibinfo{person}{M.~E.
  Delamaro}, {and} \bibinfo{person}{L. Baresi}.}
  \bibinfo{year}{2013}\natexlab{}.
\newblock \showarticletitle{Specifying automated oracles for Simulink models}.
  In \bibinfo{booktitle}{\emph{International Conference on Embedded and
  Real-Time Computing Systems and Applications}}. IEEE,
  \bibinfo{pages}{330--333}.
\newblock


\bibitem[\protect\citeauthoryear{Newton}{Newton}{1774}]
        {newton1774methodus}
\bibfield{author}{\bibinfo{person}{Isaac Newton}.}
  \bibinfo{year}{1774}\natexlab{}.
\newblock \showarticletitle{Methodus fluxionum et seriarum infinitarum}.
\newblock \bibinfo{journal}{\emph{Opuscula mathematica, philosophica et
  philologica}}  \bibinfo{volume}{1} (\bibinfo{year}{1774}).
\newblock


\bibitem[\protect\citeauthoryear{Oliveira, Kanewala, and Nardi}{Oliveira
  et~al\mbox{.}}{2014}]
        {oliveira2014automated}
\bibfield{author}{\bibinfo{person}{Rafael~AP Oliveira}, \bibinfo{person}{Upulee
  Kanewala}, {and} \bibinfo{person}{Paulo~A Nardi}.}
  \bibinfo{year}{2014}\natexlab{}.
\newblock \showarticletitle{Automated test oracles: State of the art,
  taxonomies, and trends}.
\newblock In \bibinfo{booktitle}{\emph{Advances in computers}}.
  Vol.~\bibinfo{volume}{95}. \bibinfo{publisher}{Elsevier},
  \bibinfo{pages}{113--199}.
\newblock


\bibitem[\protect\citeauthoryear{Pezze and Zhang}{Pezze and Zhang}{2014}]
        {pezze2014automated}
\bibfield{author}{\bibinfo{person}{Mauro Pezze} {and} \bibinfo{person}{Cheng
  Zhang}.} \bibinfo{year}{2014}\natexlab{}.
\newblock \showarticletitle{Automated test oracles: A survey}.
\newblock In \bibinfo{booktitle}{\emph{Advances in Computers}}.
  Vol.~\bibinfo{volume}{95}. \bibinfo{publisher}{Elsevier},
  \bibinfo{pages}{1--48}.
\newblock


\bibitem[\protect\citeauthoryear{Pill and Wotawa}{Pill and Wotawa}{2018}]
        {pill2018automated}
\bibfield{author}{\bibinfo{person}{Ingo Pill} {and} \bibinfo{person}{Franz
  Wotawa}.} \bibinfo{year}{2018}\natexlab{}.
\newblock \showarticletitle{Automated generation of (F) LTL oracles for testing
  and debugging}.
\newblock \bibinfo{journal}{\emph{Journal of Systems and Software}}
  \bibinfo{volume}{139} (\bibinfo{year}{2018}), \bibinfo{pages}{124--141}.
\newblock


\bibitem[\protect\citeauthoryear{Platzer}{Platzer}{2018}]
        {uncertainty}
\bibfield{author}{\bibinfo{person}{Andr{\'e} Platzer}.}
  \bibinfo{year}{2018}\natexlab{}.
\newblock \bibinfo{title}{Foundations of Cyber-Physical Systems}.
\newblock
\newblock


\bibitem[\protect\citeauthoryear{Richardson, Aha, and O'Malley}{Richardson
  et~al\mbox{.}}{1992}]
        {Richardson:1992:STO:143062.143100}
\bibfield{author}{\bibinfo{person}{Debra~J. Richardson},
  \bibinfo{person}{Stephanie~Leif Aha}, {and} \bibinfo{person}{T.~Owen
  O'Malley}.} \bibinfo{year}{1992}\natexlab{}.
\newblock \showarticletitle{Specification-based Test Oracles for Reactive
  Systems}. In \bibinfo{booktitle}{\emph{International Conference on Software
  Engineering}} \emph{(\bibinfo{series}{ICSE})}. \bibinfo{publisher}{ACM}.
\newblock


\bibitem[\protect\citeauthoryear{Roy and Shankar}{Roy and Shankar}{2011}]
        {roy2011simcheck}
\bibfield{author}{\bibinfo{person}{Pritam Roy} {and} \bibinfo{person}{Natarajan
  Shankar}.} \bibinfo{year}{2011}\natexlab{}.
\newblock \showarticletitle{SimCheck: a contract type system for Simulink}.
\newblock \bibinfo{journal}{\emph{Innovations in Systems and Software
  Engineering}} \bibinfo{volume}{7}, \bibinfo{number}{2}
  (\bibinfo{year}{2011}), \bibinfo{pages}{73--83}.
\newblock


\bibitem[\protect\citeauthoryear{Satpathy, Butler, Leuschel, and
  Ramesh}{Satpathy et~al\mbox{.}}{2007}]
        {10.1007/978-3-540-73770-4_6}
\bibfield{author}{\bibinfo{person}{Manoranjan Satpathy},
  \bibinfo{person}{Michael Butler}, \bibinfo{person}{Michael Leuschel}, {and}
  \bibinfo{person}{S. Ramesh}.} \bibinfo{year}{2007}\natexlab{}.
\newblock \showarticletitle{Automatic Testing from Formal Specifications}. In
  \bibinfo{booktitle}{\emph{Tests and Proofs}},
  \bibfield{editor}{\bibinfo{person}{Yuri Gurevich} {and}
  \bibinfo{person}{Bertrand Meyer}} (Eds.). \bibinfo{publisher}{Springer Berlin
  Heidelberg}, \bibinfo{address}{Berlin, Heidelberg}, \bibinfo{pages}{95--113}.
\newblock
\showISBNx{978-3-540-73770-4}


\bibitem[\protect\citeauthoryear{Schneider, Basin, Brix, Krsti{\'{c}}, and
  Traytel}{Schneider et~al\mbox{.}}{2018}]
        {10.1007/978-3-030-03769-7_20}
\bibfield{author}{\bibinfo{person}{Joshua Schneider}, \bibinfo{person}{David
  Basin}, \bibinfo{person}{Frederik Brix}, \bibinfo{person}{Sr{\dj}an
  Krsti{\'{c}}}, {and} \bibinfo{person}{Dmitriy Traytel}.}
  \bibinfo{year}{2018}\natexlab{}.
\newblock \showarticletitle{Scalable Online First-Order Monitoring}. In
  \bibinfo{booktitle}{\emph{Runtime Verification}},
  \bibfield{editor}{\bibinfo{person}{Christian Colombo} {and}
  \bibinfo{person}{Martin Leucker}} (Eds.). \bibinfo{publisher}{Springer},
  \bibinfo{address}{Cham}, \bibinfo{pages}{353--371}.
\newblock


\bibitem[\protect\citeauthoryear{Silvetti, Nenzi, Bartocci, and
  Bortolussi}{Silvetti et~al\mbox{.}}{2018}]
        {silvetti2018signal}
\bibfield{author}{\bibinfo{person}{Simone Silvetti}, \bibinfo{person}{Laura
  Nenzi}, \bibinfo{person}{Ezio Bartocci}, {and} \bibinfo{person}{Luca
  Bortolussi}.} \bibinfo{year}{2018}\natexlab{}.
\newblock \showarticletitle{Signal Convolution Logic}.
\newblock \bibinfo{journal}{\emph{arXiv preprint arXiv:1806.00238}}
  (\bibinfo{year}{2018}).
\newblock


\bibitem[\protect\citeauthoryear{Srinivasan and Leveson}{Srinivasan and
  Leveson}{2002}]
        {1067980}
\bibfield{author}{\bibinfo{person}{J. Srinivasan} {and} \bibinfo{person}{N.
  Leveson}.} \bibinfo{year}{2002}\natexlab{}.
\newblock \showarticletitle{Automated testing from specifications}. In
  \bibinfo{booktitle}{\emph{Digital Avionics Systems Conference}}. IEEE,
  \bibinfo{pages}{6A2--6A2}.
\newblock


\bibitem[\protect\citeauthoryear{Stocks and Carrington}{Stocks and
  Carrington}{1993}]
        {Stocks:1993:TTS:257572.257664}
\bibfield{author}{\bibinfo{person}{P.~A. Stocks} {and} \bibinfo{person}{D.~A.
  Carrington}.} \bibinfo{year}{1993}\natexlab{}.
\newblock \showarticletitle{Test Templates: A Specification-based Testing
  Framework}. In \bibinfo{booktitle}{\emph{International Conference on Software
  Engineering}} \emph{(\bibinfo{series}{ICSE})}. \bibinfo{publisher}{IEEE}.
\newblock


\bibitem[\protect\citeauthoryear{Watanabe, Kang, Lin, and Shiraishi}{Watanabe
  et~al\mbox{.}}{2018}]
        {Watanabe:2018:RMS:3195970.3199856}
\bibfield{author}{\bibinfo{person}{Kosuke Watanabe}, \bibinfo{person}{Eunsuk
  Kang}, \bibinfo{person}{Chung-Wei Lin}, {and} \bibinfo{person}{Shinichi
  Shiraishi}.} \bibinfo{year}{2018}\natexlab{}.
\newblock \showarticletitle{Runtime Monitoring for Safety of Intelligent
  Vehicles}. In \bibinfo{booktitle}{\emph{Annual Design Automation Conference}}
  \emph{(\bibinfo{series}{DAC})}. \bibinfo{publisher}{ACM}.
\newblock


\bibitem[\protect\citeauthoryear{Whittle, Sawyer, Bencomo, Cheng, and
  Bruel}{Whittle et~al\mbox{.}}{2009}]
        {whittle2009relax}
\bibfield{author}{\bibinfo{person}{Jon Whittle}, \bibinfo{person}{Pete Sawyer},
  \bibinfo{person}{Nelly Bencomo}, \bibinfo{person}{Betty~HC Cheng}, {and}
  \bibinfo{person}{Jean-Michel Bruel}.} \bibinfo{year}{2009}\natexlab{}.
\newblock \showarticletitle{Relax: Incorporating uncertainty into the
  specification of self-adaptive systems}. In
  \bibinfo{booktitle}{\emph{Requirements Engineering Conference}}. IEEE,
  \bibinfo{pages}{79--88}.
\newblock


\bibitem[\protect\citeauthoryear{Zander, Schieferdecker, and Mosterman}{Zander
  et~al\mbox{.}}{2012}]
        {zander:12}
\bibfield{author}{\bibinfo{person}{Justyna Zander}, \bibinfo{person}{Ina
  Schieferdecker}, {and} \bibinfo{person}{Pieter~J Mosterman}.}
  \bibinfo{year}{2012}\natexlab{}.
\newblock \bibinfo{booktitle}{\emph{Model-based testing for embedded systems}}.
\newblock \bibinfo{publisher}{CRC Press}.
\newblock


\bibitem[\protect\citeauthoryear{Zheng, Julien, Kim, and Khurshid}{Zheng
  et~al\mbox{.}}{2017}]
        {zheng2017perceptions}
\bibfield{author}{\bibinfo{person}{Xi Zheng}, \bibinfo{person}{Christine
  Julien}, \bibinfo{person}{Miryung Kim}, {and} \bibinfo{person}{Sarfraz
  Khurshid}.} \bibinfo{year}{2017}\natexlab{}.
\newblock \showarticletitle{Perceptions on the state of the art in verification
  and validation in cyber-physical systems}.
\newblock \bibinfo{journal}{\emph{Systems Journal}} \bibinfo{volume}{11},
  \bibinfo{number}{4} (\bibinfo{year}{2017}), \bibinfo{pages}{2614--2627}.
\newblock


\end{thebibliography}

\renewcommand{\thesubsection}{\Alph{subsection}}
\section*{Appendix}
Section~\ref{sec:rfol2stl} describes how an STL formula can be converted into an equivalent RFOL formula.
Section~\ref{sec:proofs} contains the proofs for the Theorems of Section~\ref{sec:contribution}

\subsection{Transforming STL formulae in RFOL}
\label{sec:rfol2stl}
We show that any STL formula can be expressed in RFOL. 
Let $x$ be a real variable and $c \in \real$, an STL formula can be described using the following grammar:
\begin{align}
& \phi && ::= &&  x \sim c \mid \neg \phi \mid \phi_1 \lor \phi_2 \mid \phi_1 \wedge \phi_2 \mid \LTLf_{[a,b]} \phi \mid \LTLg_{[a,b]} \phi & \nonumber \\
& & &&  &  \mid  \phi_1 \LTLu_{[a,b]} \phi_2 \mid   \phi_1 \LTLr_{[a,b]} \phi_2 \nonumber&
\end{align}
where $x \in X$, $\sim \in \{ <, \leq \}$.
$\neg$, $\wedge$ $\lor$ are boolean operators while 
$\LTLu_{[a,b]}$, $ \LTLf_{[a,b]}$ are temporal operators.

Any STL can be expressed in RFOL by executing a procedure that follows two steps.

\texttt{Step 1.} The STL formula is rewritten into an equivalent STL formula by executing the following operations:
\begin{itemize}
\item compute the negation normal form (NNF) of the STL formula;\footnote{A formula is in negation formal form 
if negation $\neg$ occurs only directly in front of atoms.} 
\item transform any atom in the form $\neg (x < c)$ into $x \geq c$ and  $\neg (x \leq c)$ into $x > c$. 
\end{itemize}
Note that the previous procedure pushes all the negations into formula atoms.

\texttt{Step 2.} The syntax tree of the STL formula is analyzed from the root to the leaves and every operator of the resulting STL formula is transformed into RFOL as follows: 

\begin{itemize}
\item every STL subformula $\varphi$  in the form $x \sim c$, $\phi_1 \lor \phi_2$ and $\phi_1 \wedge \phi_2$ is converted into an equivalent RFOL subformula that uses the same boolean or relational operator;
\item every STL subformula $\varphi$ in the form $\LTLf_{[a,b]} \phi$ is converted into
\begin{enumerate}
\item the RFOL formula $\exists t \in [a,b]: \phi$ if $\varphi$ is \emph{not} nested into another  formula that uses any temporal operators;
\item the RFOL formula $\exists t \in [t_f+a,t_f+b]: \phi$ if $\varphi$ is nested into another  formula that uses temporal operators. The timed variable $t_f$ is the timed variable introduced in the RFOL formula when the other formula is mapped into RFOL;
\end{enumerate}
\item every STL subformula $\varphi$ in the form $\phi_1 \LTLu_{[a,b]} \phi_2$ is converted into 
\begin{enumerate}
\item the RFOL formula $\exists t \in [a,b]: (\phi_2 \wedge (\forall t^\prime \in [a,t]: \phi_1))$ if $\varphi$ is \emph{not} nested into another  formula that uses the temporal operators;
\item the RFOL formula $\exists t \in [t_f+a,t_f+b]: (\phi_2 \wedge (\forall t^\prime \in [t_f+a,t]: \phi_1))$ if $\varphi$  is nested into another  formula that uses temporal operators. The timed variable $t_f$ is the timed variable introduced in the RFOL formula when the other formula is mapped into RFOL;
\end{enumerate} 
\item every STL subformula $\varphi$ in the form $\LTLg_{[a,b]} \phi$ is converted into
\begin{enumerate}
\item the RFOL formula $\forall t \in [a,b]: \phi$ if  $\varphi$ is \emph{not} nested into another  formula that uses temporal operators;
\item the RFOL formula $\forall t \in [t_f+a,t_f+b]: \phi$ if $\varphi$ is nested into another  formula that uses temporal operators. The timed variable $t_f$ is the timed variable introduced in the RFOL formula when the other formula is mapped into RFOL;
\end{enumerate}
\item every STL subformula $\varphi$ in the form $\phi_1 \LTLr_{[a,b]} \phi_2$ is converted into 
\begin{enumerate}
\item the RFOL formula  $\exists t \in [a,b]: ((\phi_2 \wedge \phi_1) \wedge (\forall t^\prime \in [a,t]: \phi_2)) \vee \forall t \in [a,b]: (\phi_2)$ if the formula $\varphi$ is \emph{not} nested into another  formula that uses the temporal operators;
\item the RFOL formula $\exists t \in [t_f+a,t_f+b]: ((\phi_2 \wedge \phi_1) \wedge (\forall t^\prime \in [t_f+a,t]: \phi_2)) \vee \forall t \in [t_f+a,t_f+b]: \phi_2$ if $\varphi$  is nested into another  formula that uses temporal operators. The timed variable $t_f$ is the timed variable introduced in the RFOL formula when the other formula is mapped into RFOL.
\end{enumerate} 
\end{itemize}
Note that it is trivial to show correctness of the proposed encoding as the generated RFOL formulae are a direct encoding of the semantics of 
the corresponding STL formulae.

\subsection{Proofs}
\label{sec:proofs}
We provide a sketch of the proofs for the theorems presented in Section~\ref{sec:contribution}.
\theoremshifting
\begin{proof}[Proof Sketch]
The proof follows from the fact that every time interval and every time variable are shifted consistently together, i.e.,
if the bound of a variable $t$ is changed from $\langle n_1, n_2 \rangle$ into $\langle n_1+d, n_2+d \rangle$ the occurrences of 
$t$ in the formula are replaced with $t-d$ ensuring that the correct value of signal is considered in the evaluation of the formula.
Hence, the semantics of formulas is not impacted.   
\end{proof}

\theoremtranslation

\begin{proof}[Proof Sketch]
The proof follows by structural induction on \ourlogic\ formulas.
The  atomic terms are constructed by rules 1, 2 and 4. 
It is easy to show that these rules correctly compute the semantics of  these atomic terms, 
and further, they do not rely on a signal value that is not yet generated up to the current time $t$. 
The inductive proof for rules 3, 5, 6, 7, 8, and  9 is as follows: (1)~Each rule correctly computes the value of \ourlogic\ constructs based on the \ourlogic\ semantics, and (2) the rules do not rely on a signal value that is not yet generated  up to the current time $t$. In particular, for Rule8, due to the interval-shifting procedure, we know that the value of $\phi$ in 
$\forall t \in \langle \tau_1, \tau_2 \rangle \colon  \phi$ at time $t$ is correctly computed inductively. 
Finally, by Theorem~\ref{thm:shifting}, we know that our time and interval shifting procedures are semantic preserving. Therefore, our translation for $\varphi_\Uparrow$, i.e., $M_\varphi$, is able to correctly compute the semantics of $\varphi$.  
\end{proof}

\theoremstop
\begin{proof}[Proof Sketch]
By Definition~\ref{def:rfol} $\varphi$ is closed. 
Thus, the output of the oracle is the output of a block that is generated using either Rule6, Rule7 or Rule8.
Indeed, as there is no free variable, i.e., any variable is scoped by existential and universal quantifiers, the output signal of a block generated using Rule9 cannot be the output of the oracle as it has to be then processed by at least a block generated by Rule8.
The output signal generated from the Simulink blocks obtained from the univeral quantifiers by applying Rule 8 is decreasing as at each simulation step they compute the minimum of a previously computed value and a new value which is  less than or equal to the previously computed value.
Furthermore, if $t>d$, the output signal generated from the Simulink blocks obtained from the existential quantifiers by applying Rule 8 is constant by construction.
Finally, the maximum and minimum of two decreasing signals is also decreasing.
Thus, each $e_i \in \{\{e_1\}, \{e_2\}, \ldots, \{e_k\}\}$  is decreasing over the time interval  
$(d,t_u]$. 
\end{proof}

\end{document}